\documentclass[journal,compsoc]{IEEEtran}
\usepackage{multirow}
\usepackage{blindtext, graphicx}
\usepackage{listings}
\usepackage{hyperref}

\usepackage{amssymb}
\usepackage{graphicx}
\usepackage{booktabs}
\usepackage{graphicx}

\usepackage{multirow}
\usepackage[dvipsnames]{xcolor}
\usepackage{fancyhdr}
\usepackage{enumitem}

\usepackage{amsmath}
\usepackage{amsfonts}
\usepackage{algorithmic}
\usepackage{algorithm}
\usepackage{array}

\usepackage{textcomp}
\usepackage{stfloats}
\usepackage{url}
\usepackage{verbatim}
\usepackage{cite}

\usepackage[utf8]{inputenc}
\newtheorem{definition}{Definition}
\newtheorem{observation}{Observation}
\newtheorem{theorem}{Theorem}
\newtheorem{lemma}{Lemma}

\newcommand{\proof}[1]{\vspace{+3pt}\noindent\textbf{\textit{{Proof.}}}}

\newcommand{\citet}{\cite}
\newcommand{\citep}{\cite}
\newcommand{\qed}{$\square$}

\usepackage{amsmath}

\begin{document}

\title{Micro Operational Robustness of \\LLMs on Code Generation}

\author{Debalina Ghosh Paul, 
Hong Zhu*\thanks{*Corresponding author, email: hzhu@brookes.ac.uk}, IEEE Senior Member, 
Ian Bayley
\thanks{The authors are with the School of Engineering, Computing and Mathematics, Oxford Brookes University, Oxford, OX3 0BP, United Kingdom. 
}
\thanks{Last revision on \today}
}

\markboth{Micro Operational Robustness of Code Generation}%
{Paul, D., Zhu, H. and Bayley, I.}

\maketitle

\begin{abstract}
It is now common practice in software development for large language models (LLMs) to be used to generate program code. It is desirable to evaluate the robustness of LLMs for this usage.  
This paper is concerned in particular with how sensitive LLMs are to variations in descriptions of the coding tasks. However, existing techniques for evaluating this robustness are unsuitable for code generation because the input data space of natural language (NL) descriptions is discrete and sparse.
To address this problem, we propose a robustness evaluation method called \emph{scenario domain analysis}, which aims to find the expected minimal change in the NL descriptions of coding tasks that would cause the LLMs to produce incorrect outputs.
We have formally proved the theoretical properties of the method and also conducted extensive experiments to evaluate the robustness of four state-of-the-art LLMs: \emph{Gemini-pro},  \emph{Codex},  \emph{Llamma2} and  \emph{Falcon 7B}, and have found that we are able to rank these with confidence from best to worst.  
Moreover, we have also studied how robustness varies in different scenarios, including the variations with the topic of the coding task and with the complexity of its sample solution, and found that robustness is lower for more complex tasks and also lower for more advanced topics, such as multi-threading and data structures.  
\end{abstract}


\section{Introduction}\label{sec:Introduction}

Robustness has been a hard problem and an active topic of research in the field of machine learning (ML) ever since Szegedy \emph{et al.}'s discovery of the universal existences of adversarial examples in deep neural networks (DNNs) \citet{szegedy2014}. In general terms, the robustness of an artificial intelligence (AI) system, as defined by ISO Standard 24029 \citet{iso24029-1-2023}, is \emph{the ability of the system to maintain its level of performance under any circumstances}. It is a critical quality attribute of ML applications and existing works can be classified into two types, according to the circumstances concerned: 

\begin{itemize}
\item \emph{Adversarial robustness}. Used when an ML model, or its application system, is under attack from an adversary who intends to force the ML model to produce incorrect outputs. Robustness, in this context, means that the ML model performs as expected by resisting such adversarial attacks. It is closely related to security and has been an active research topic for the past decade; see \citep{khamaiseh2022,tekeste2026adversarial} for recent surveys. 

\item \emph{Operational robustness}. Used when the operational conditions (or scenarios) vary naturally without malicious attacks. Here, robustness is the ability of the ML model to maintain its performance in a wide spectrum of scenarios. For example, an autonomous vehicle should drive safely not only in fine weather but also in rain, snow and fog.
\end{itemize}

This paper is concerned with the operational robustness of large language models (LLMs) as code generation tools, which turn NL descriptions of requirements into program code. This is a widely used application of LLMs and there is overwhelming evidence that it can significantly improve the productivity of software engineers although concerns have been raised regarding the quality of LLM-generated code; see, for example, recent surveys \citet{DORA2024} and \citet{Harding2025}. Operational robustness affects the usability of these code generation tools and therefore it needs to be understood.

There are two approaches to assessing the robustness of an ML model.

\begin{itemize}
\item The \emph{micro approach} focuses on the ML model's sensitivity to changes in the input, and evaluates the robustness by analysing the impact of changes on each input value. For example, if the ML model is the object detection model for an autonomous vehicle, a micro robustness property is the expected minimal number of pixels in an image to be changed to cause error in object detection. This is the approach taken in most research on adversarial robustness.
\item The \emph{macro approach} assesses the stability in performance on various sub-domains of the input space, each of which represents a different scenario. For instance, in the autonomous vehicles example given above, a macro robustness property is that the accuracy of object detection remains high (e.g. over 90\%) in all weather conditions. This approach is used for measuring operational robustness. 
\end{itemize}

Our previous work \citep{ghosh2024} found a significant decrease in the correctness of code produced by LLMs when the complexity of the coding task increases. Similarly, in \citep{CodeSmell2025}, we found that the presence of code smells, another indicator of code quality, varies significantly according to coding topics and complexity of the task. 
All these are macro robustness properties. 

However, far less is known about micro operational robustness. For example, when a software developer entering a prompt to generate code, how sensitive is the LLM to the choice of words in the prompt? That is, can replacing one word with its synonym in a coding task description lead to a completely different program? Similarly, when a prompt does not result in the required code, how many words in the prompt would the developer need to change to achieve his/her goal? This is the problem to be addressed in this paper. 

Unfortunately, there is a lack of effective and efficient techniques and methods for assessing such micro operational robustness of ML models. In this paper, we propose a new method for this called \emph{Scenario Domain Analysis} and we develop a new scenario domain exploration algorithm that is suitable for generative tasks on discrete and sparse input domain like code generation. 

The remainder of the paper is organised as follows.

Section \ref{sec:ProposedMethod} sets the theoretical foundation of the proposed method by formally defining the notion of micro operational robustness in a given scenario, studying the properties of discrete and sparse input domains, presenting an algorithm that is suitable for such input domains. 
In Section \ref{sec:ApplyToCodeGen}, this theoretical framework is applied to code generation as an example.
Section \ref{sec:Experiment} reports the experiments on LLMs to demonstrate the effectiveness and efficiency of the proposed method. 
We will also discuss the threats to the validity of the experiment.
Section \ref{sec:RelatedWork} compares with related works and summarises the main contributions of this paper. 
Finally, Section \ref{sec:Conclusion} concludes the paper with a summary of the main contributions and a discussion of possible directions for future work. 

\section{The Proposed Method}\label{sec:ProposedMethod}

This section presents the proposed method called \emph{scenario domain analysis} for testing and evaluating micro operational robustness of ML models. We first formally define the notion of micro operational robustness and study its mathematical properties. Then, we present an algorithm for evaluating the micro operational robustness in a scenario that is suitable for discrete and sparse scenario domains. Finally, the correctness of the algorithm will be formally analysed so that the underlying assumptions of the algorithm are identified to provide practical guidelines for valid application of the algorithm. The proofs of the theorems and lemmas can be found in the Appendix. In the next section, the proposed algorithm will be illustrated and instantiated with technical details to the problem of code generation. 

\subsection{Definition of Micro Operational Robustness} \label{sec:TheoreticalFramework}

The basic idea of the proposed method is, for each test case in a benchmark, to systematically explore the input space in the sub-domain of the operational scenario surrounding the test case to discover the minimal deviation from the test case in the operation scenario that causes the ML model fail and the maximal size of the safe zone in which the ML model will not fail. This is analogous to evaluating adversarial robustness, in which the minimal perturbation on the input that causes the ML model to fail is sought after \cite{moosavi2016}. In continuous input spaces like images, the minimal failure causing deviation is the same as the maximal safe size. However, for discrete and sparse input spaces like NLP, there may be a gap between them. Thus, our definition of micro operational robustness in an operational scenario use the expected values for both quantities over the data distribution to indicate how sensitive the ML model is to the changes to the input. 

Generally speaking, a scenario in the use of an ML model $M$ is a specific subset of input data defined by a specific operational condition on which the user invokes the ML model to achieve a specific usage purpose \citep{zhu2019a,zhu2023scenario,ghosh2024}. A scenario $s$, therefore, can be characterised by a sub-domain $D^s$ of the input space $D$ of the ML model $M$ that satisfies the operational condition. 

For example, driving in rain is a scenario in the operation of an autonomous vehicle. For the object detection ML model of the autonomous vehicle, this scenario can be characterised by the set of images of road traffic that satisfy the raining weather condition. 

Another example of a scenario in using a LLM as a coding tool is to develop a database component as the back-end of a SaaS application. This scenario can be characterised by the set of prompts that describe coding tasks to generate relational database SQL statements. 

As discussed in Section \ref{sec:Introduction}, micro operational robustness is concerned with how sensitive a ML model is to variations in the input data. Therefore, we assume that there is a distance metric (or similarity metric) $\|\cdot , \cdot \|$ defined on the input data space $D$ to measure the degree of variation. 

Moreover, we also assume that there is a test oracle that can determine if the ML model $M$ behaves differently between two input data $x$ and $y$. Considering $x$ as a deviation from a test case $t$, we say that ML model $M$ fails on $x$ if $M$ behaves sufficiently differently according to the test oracle on $x$ from its behaviour on $t$, and we write $Fail^M_t(x)$. 

For example, for a classification ML model $M$, $Fail^M_t(x)$ could be simply defined as $M(t) \neq M(x)$. For a generative ML model $M$, a possible definition of $Fail^M_t(x)$ could be $\|M(t), M(x)\| \geq \delta$ for a given real value $\delta > 0$, where $\|\cdot, \cdot\|$ is a metric on the co-domain of $M$. 

For the ML model to be robust in operation, one would expect that it resists small variations in its input; that is, it will behave in the same way when the input is changed slightly. For example, one would expect an autonomous vehicle to drive the same way in drizzle as in fine weather. Inspired by Moosavi \emph{et al.}'s notion of the minimal perturbation \citet{moosavi2016}, we now define the notion of \emph{minimal deviation} that causes failure in a scenario $s$. 

\begin{definition}(Minimal distance to failure in a scenario)\label{def:MinimalDistanceToFailure}

Let $t \in D$ and $D^s \subset D$ be the subset of the input space representing an operational scenario $s$.
 
An ML model $M$'s \emph{minimal distance from $t$ to failure} in the scenario $s$, denoted by $\Delta_{D^s}(t,M)$, is defined as follows. 

\begin{equation}
\Delta_{D^s}(t,M)=\min_{x \in D^s}\{\|x,t\| ~|~ Fails^M_t(x)\}. 
\end{equation}
The data $x\in D^s$ such that $\|x,t\|=\Delta_{D^s}(t,M)$ is called a \emph{minimal deviation} of $t$ in scenario $s$. 
\qed
\end{definition}

By the definition of $\Delta_{D^s}(t,M)$, we have that the following lemma. 

\begin{lemma}\label{lmm:lmmA}
For all $x \in D^s$, we have that 
\[\left(\|x,t\| < \Delta_{D^s}(t,M) \Rightarrow \neg Fails^M_t(x)\right). ~~ \square \] 
\end{lemma}

Informally, Lemma \ref{lmm:lmmA} states that in the ML model $M$ does fail on any input within the distances $\Delta_{D^s}(t,M)$ from $t$. This is equivalent to the following. 
\[\forall x \in D^s. \left(Fail^M_t(x) \Rightarrow \|x,t\| \geq \Delta_{D^s}(t,M)\right).\]

In other words, $\Delta_{D^s}(t,M)$ defines a ``safe zone'' in which the ML model will also behave the same way as $t$. Outside the safe zone, there is no guarantee that the ML model will behave the same way as on the test case $t$, although that does not mean $M$ will always fail. The data points in the safe zone can be considered to be ``safe''.  

The notion of \emph{safe} can be formally defined as follows. 

\begin{definition}(Safe mutant)\label{def:Safe}

Let $t \in D$. We say that $x \in D^s$ is \emph{safe} for model $M$ w.r.t. to $t$, denoted by $Safe^M_t(x)$, if and only if (a) the model $M$ does not fail on $x$, and (b) $M$ does not fail on any points $x'$ in $D^s$ whose distance to $t$ is less than the distance from $x$ to $t$. Formally, 
\begin{eqnarray}
\lefteqn{Safe^M_t(x) = \neg Fails^M_t(x)~ \wedge }\nonumber \\
&& \forall x' \in D^s.( \| x', t\| < \| x,t\| \Rightarrow \neg Fails^M_t(x'))
\end{eqnarray}
\qed
\end{definition}

Utilising the notion of safe data point, we can define the notion of maximal safe distance in a scenario as follows. 

\begin{definition}(Maximal safe distance in a scenario)\label{def:SafeDistance}

The \emph{maximal safe distance} from $t$ in a scenario domain $D^s$ for ML model $M$, denoted by $\nabla_{D^s}(t,M)$, is defined as follows. 
\begin{equation}
\nabla_{D^s}(t,M)=\max_{x \in D^s}\{\|x,t\| ~| Safe^M_t(x)\}. 
\end{equation}
\qed
\end{definition}

\begin{figure}[hb]
\centering
\includegraphics[width=8.5 cm]{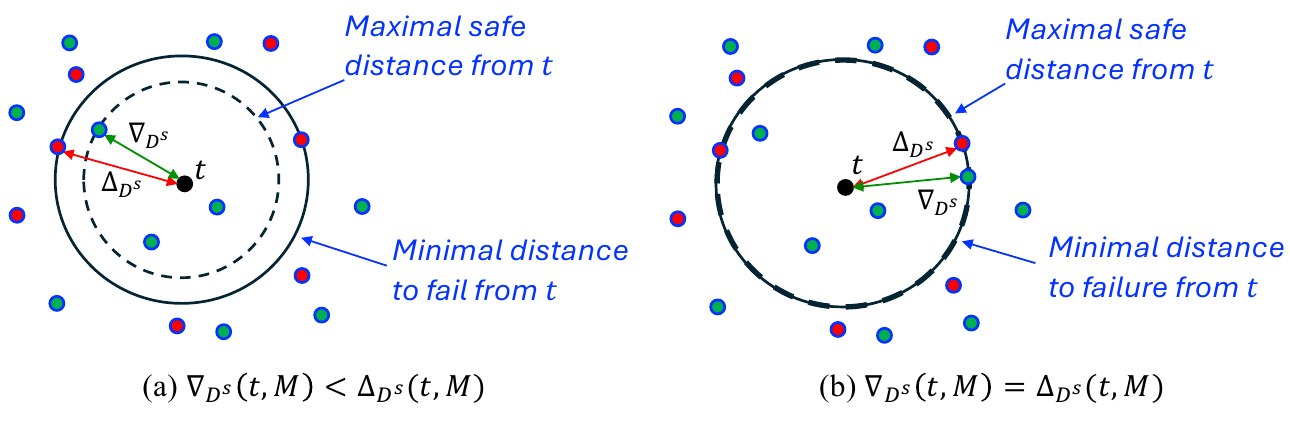}
\caption{Illustration of The Notions of Minimal Distance to Failure and Maximal Safe Distance} 
\label{fig:illustrationOfNotions}
\end{figure}

Fig. \ref{fig:illustrationOfNotions} illustrates the notions of maximal distance to failure and minimal safe distance from a test case. In both Fig. \ref{fig:illustrationOfNotions}(a) and (b), the center point marked as $t$ is the test case. The dots surrounding $t$ are the input data points in the scenario domain $D^s$ nearby to the test case $t$. The green dots are those inputs $x$ on which the ML model $M$ behaves the same way as on $t$, while the red dots are those inputs $x$ on which ML model $M$ behaves differently, i.e. $Fail^M_t(x)$. Those green dots within or on the dashed line circle are safe, but all the other dots are not safe. The red and green arrows indicate  the minimal distance to failure and the maximal safe distance, respectively. 

From Definition \ref{def:Safe} and \ref{def:SafeDistance}, it is easy to prove the following lemmas. 

\begin{lemma}\label{lmm:lmmC}
$\forall x \in D^s.\left(\|x,t\| < \nabla_{D^s}(t,M) \Rightarrow Safe^M_t(x)\right)$. \qed
\end{lemma}

\begin{lemma}\label{lmm:lmmD}
$\forall x \in D^s.\left(Safe^M_t(x) \Rightarrow \|x,t\| \leq \nabla_{D^s}(t,M)\right)$. \qed
\end{lemma}

Lemma \ref{lmm:lmmD} is equivalent to the following. 
\[\forall x \in D^s.\left(\|x,t\| > \nabla_{D^s}(t,M) \Rightarrow \neg Safe^M_t(x)\right).\] 

This means, for all data points $x$ in $D^s$ whose distance from $t$ is greater than $\nabla_{D^s}(t,M)$, $x$ is not safe. Meanwhile, Lemma \ref{lmm:lmmC} states that, for all $x$ in $D^s$ whose distance from $t$ is less than $\nabla_{D^s}(t,M)$, we have that $x$ is safe. 

However, for an $x$ in $D^s$ whose distance from $t$ is equal to $\nabla_{D^s}(t,M)$, the definition of $\nabla_{D^s}(t,M)$ does not imply whether $x$ is safe or not. In fact, in a discrete and sparse scenario domain $D^s$, by Definition \ref{def:SafeDistance}, there exists an input $x$ in $D^s$ such that $\|x,t\| = \nabla_{D^s}(x,M)$ and $Safe^M_t(x)$.\footnote{A data space $D$ is defined as \emph{discrete and sparse}, if and only if for all real number $\delta >0$ and all $t \in D$, there is only a finite number of $x \in D$ such that $\|x, t\| \leq \delta$.} 
On the other hand, it is also possible that there exists an $x$ in $D^s$ such that $\|x,t\| = \nabla_{D^s}(x,M)$ but $Fail^M_t(x)$, thus $\neg Safe^M_t(x)$. For example, Fig \ref{fig:illustrationOfNotions}(b) shows the situation when there are two input data of the same distance to $t$ that the ML model is safe on one but fails on the other. 

From Definition \ref{def:MinimalDistanceToFailure} and \ref{def:SafeDistance}, we can also prove the following relationship between minimal distance to failure and maximal safe distance. 

\begin{theorem}\label{thm:thm1}
$\forall t \in D. \left(\nabla_{D^s}(t,M) \leq \Delta_{D^s}(t,M)\right)$. 
\qed
\end{theorem}

Fig. \ref{fig:illustrationOfNotions}(a) shows the situation when the maximal safe distance $\nabla_{D^s}(t,M)$ is less than the minimal distance to failure $\Delta_{D^s}(t,M)$ , while Fig. \ref{fig:illustrationOfNotions}(b) is when they are equal.  

Note that the equality in the above equation always holds for continuous or dense input domains. However, for discrete and sparse input spaces, there may be a gap between the maximal safe distance and minimal distance to failure as shown in Fig \ref{fig:illustrationOfNotions}(a). Therefore, in continuous or dense discrete inputs domains, it is sufficient to use either minimal distance to failure or maximal distance to characterise robustness. However, for discrete and sparse input domains, using only one of them is insufficient. 

No matter whether or not the input space is discrete or continuous, we have the following important common property about the relationships between maximal safe distance and minimal distance to failure. 

\begin{theorem}\label{thm:NablaDelta}
There is no $x \in D^s$ such that $\nabla_{D^s}(t,M) < \|x, t\| < \Delta_{D^s}(t,M)$. \qed
\end{theorem}

Having defined the notions of minimal distance to failure and maximal safe distance, we can now formally define the notion of \emph{micro operational robustness in a scenario}. 

\begin{definition}(Micro Operational Robustness in A Scenario)

The micro operational robustness in a scenario $s$ is an ordered pair of real values $\left< \rho^{*}_s, \rho^{o}_s \right>$, which are defined by the following equations. 
\begin{eqnarray}
\rho^*_s(M) = E_{x}\left(\nabla_{D^s}(x,M)\right) \\
\label{equ:RhoStar-Metric}
\rho^{o}_s(M)= E_{x}\left(\Delta_{D^s}(x,M)\right) 
\label{equ:Rho-Metric}
\end{eqnarray}
\qed
\end{definition}

\subsection{An Exploration Strategy for Sparse Domains}\label{sec:Strategy}

We regard robustness testing and evaluation as a form of exploratory testing that aims at discovering the minimal distance to failure and the maximal safe distance from each point $t$ in a benchmark $T$. As with the exploratory testing to find the boundary of classification models studied by Zhu and Bayley  \citet{zhu2022discovering}, there may be many different strategies to explore the sub-domain $D^s_\delta(t)$ surrounding a test case $t$ in the benchmark test dataset $T$. Here, we propose a highly efficient strategy that requires a small number of invocations of the ML model when the input space is discrete and sparse such as the input domain of code generation. This strategy is represented in the form of an algorithm so that it can be implemented for automated testing and evaluation of a ML model's micro operational robustness. The algorithm is given in Alg. \ref{alg:ScenarioDomainAnalysis}. 

\begin{algorithm}[ht]
\scriptsize
\caption{Exploration of Sparse Scenario Domain}
\label{alg:ScenarioDomainAnalysis}
\begin{algorithmic}[1]
\REQUIRE $ $ \\
    $T: \mathcal{P}(D)$; \textcolor{blue}{/* Test dataset */}\\
    $M: D \rightarrow C$; \textcolor{blue}{/* The ML model under test */}\\
    $GenMuts: D \rightarrow \mathcal{P}(D)$; \textcolor{blue}{/* Procedure to generate a set of mutants */}\\  
    $\|\cdot,\cdot\|: D \times D \rightarrow \mathcal{R}^+$; \textcolor{blue}{/* Distance metric on $D$ */}\\      
	$Fail: C \times C \rightarrow Bool$; \textcolor{blue}{/* Test oracle */}
\ENSURE  $ $\\
    $\left<R^*, R^o \right>: \mathcal{R}^+ \times \mathcal{R}^+$; 
\STATE $listLS: List(R^+) \leftarrow []$; 
\STATE $listFF: List(R^+) \leftarrow []$;
\FOR{each $t$ in $T$}
	\STATE $listD:List(R^+) \leftarrow []$; \textcolor{blue}{/* Distances of tested mutants to $t$ */}
	\STATE $mut \leftarrow GenMuts(t)$; \textcolor{blue}{/* The current set of mutants of $t$*/}
	\STATE $ls \leftarrow 0$; \textcolor{blue}{/* $\|t,t\|=0$ */}
	\REPEAT {
		\STATE $listMD:List(D\times R^+)\leftarrow[]$;\textcolor{blue}{/* Current mutant-distance pairs */}
		\FOR{each $x$ in $mut$}
    			\STATE $listMD \leftarrow + \left<x,\|t,x\| \right>$;
		\ENDFOR
		\STATE $Sort(ListMD)$; \textcolor{blue}{/* in ascending order of distance to $t$ */}
		\STATE $c_t \leftarrow M(t)$; \textcolor{blue}{/* Test ML model $M$ on input $t$ to generate $C_t$ */}
		\FOR{each $\left<x, d_x \right>$ in $listMD$ \textcolor{blue}{/* in ascending order of distances */}}
    			\STATE $c_x \leftarrow M(x)$ \textcolor{blue}{/* Test ML model $M$ on $x$ to generate $C_x$ */}  
   			\IF{$Fail(c_t,c_x)$}
        			\STATE $ff \leftarrow d_x$; \textbf{break};
   			\ELSE 
   				\STATE $ls \leftarrow d_x$;
    			\ENDIF
		\ENDFOR
		\IF{($ff == null$) \textcolor{blue}{/* No FF is found */}} 
			\STATE \textcolor{blue}{/* Expand the mutant set */} 
			\STATE $mut \leftarrow GenMuts(t)$; \textcolor{blue}{/* Generate more mutants */}
			\FOR {$\left<x,d_x\right> \in listMD$} 
				\STATE $insert(d_x, listD)$;\textcolor{blue}{/* Insert $d_x$ into $listD$ in ascending order */}
			\ENDFOR
		\ENDIF
	}
	\UNTIL {($ff <> null$)};
	\STATE $listFF \leftarrow + ff$; 
    \IF{($listD = []$)\textcolor{blue}{/* No expansion */}}
        	\STATE $listLS \leftarrow + ls$;
    	\ELSE        	
        	\STATE $listLS \leftarrow + Max\{ls, Max\{d_x \leq ff |d_x \in listD\}\}$;
    	\ENDIF
\ENDFOR
\STATE $R^{*} \leftarrow \frac{1}{|T|} \sum\{d_x \in listLS\}$; 
\STATE $R^{o} \leftarrow \frac{1}{|T|} \sum\{d_x \in listFF\}$; 
\RETURN $\left<R^{*}, R^{o}\right>$
\end{algorithmic}
\end{algorithm}

\begin{figure} [hbt]
\includegraphics[width=8.8 cm]{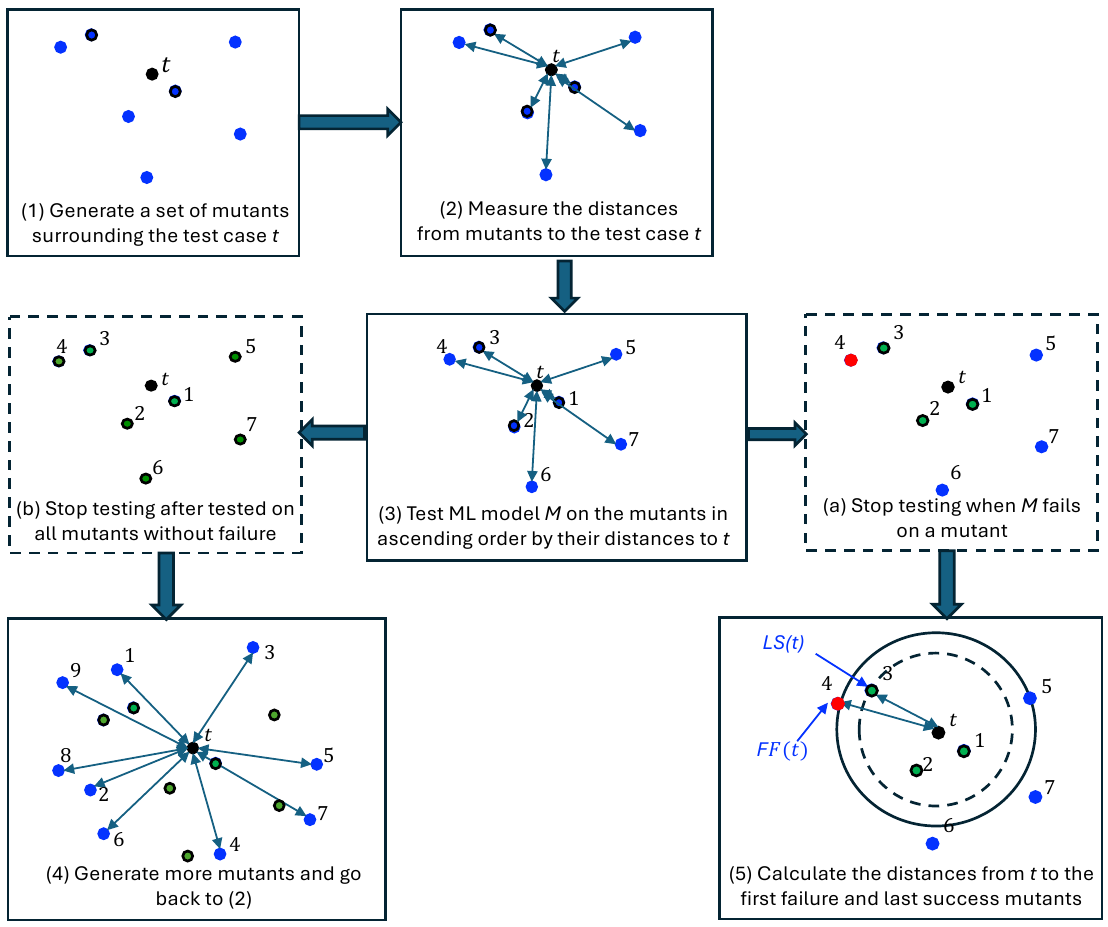}
\caption{Illustration of The Exploration Strategy for Sparse Scenario Domains}
\label{fig:IllustrationOfAlg}
\end{figure}

As detailed in Alg. \ref{alg:ScenarioDomainAnalysis} and illustrated in Fig. \ref{fig:IllustrationOfAlg}, in this strategy, the testing process consists of the following steps. 

\begin{list}{Step 1:}
\item \emph{Generating sample sets of the scenario domain.}
\end{list}

Given a test case $t$ in the benchmark test dataset $T$, the first step is to systematically generate a set $Mut_t$ of test cases in the scenario $s$ surrounding $t$; see Line 5 of Alg. \ref{alg:ScenarioDomainAnalysis}, where the set of generated test cases are saved to variable $mut$. 

The generated test cases in $Mut_t$ are called the \emph{mutants} of $t$, while the original test case $t \in T$ is called the \emph{seed} of the mutants. We require that the set $Mut_t$ of mutants are generated systematically in the sense that it is \emph{valid} and \emph{complete}. Informally, validity means that the mutants really represent the operational scenario in which the robustness of the ML model is under test. Completeness means that the mutants cover all possible variations of the inputs in the scenario. 

It is worth noting that, even if the input domain in a scenario is discrete and sparse, it could be infinite. Therefore, it is not practical to require the set of mutants be absolutely complete. Therefore, instead, we only need to generate a finite subset of the input domain that is relatively complete in the sense that it covers all variations up to a certain distance (or similarity) from the seed test case so that the minimal distance to failure and the maximal safe distance can be found in the finite set of mutants. It is possible that a set of mutants generated is not large enough even to be relatively complete. In such a case, the set of mutants will be expanded by generating more mutants; see Step 4. 

\begin{list}{Step 2:}
\item \emph{Measuring the distances from mutants to the seed.} 
\end{list}

The mutants $x$ in the sample set $Mut_t$ generated from a seed $t$ are measured for their distances to the seed $t$ using a distance metric $\|x,t\|$. These mutants are then sorted in ascending order according to their distances from the seed. See Line 8 - 12 of Alg. \ref{alg:ScenarioDomainAnalysis}. 

For all $x,y \in D$, we write $x \prec_t y$ for $\|x,t\|< \|y,t\|$, $x \approx_t y$ for $\|x,t\| = \|y,t\|$, and $x \preccurlyeq_t y$ for $\|x,t\| \leq \|y,t\|$. 

So, the sorted set $Mut_t$ of mutants can be represented in the form 
\[x_1 \preccurlyeq_t x_2 \preccurlyeq_t \cdots \preccurlyeq_t x_n, \]
where $n=|Mut_t|$. 

Note that, the metric $\|\cdot, \cdot \|$ on the input domain of the ML model $M$ measures how conceptually close the two inputs are to each other. The smaller the value, the closer they are, and if the value equals 0 then two inputs are identical. 

In practice, there may be many different distance metrics available on a given input domain. For example, on the domain of images, distance metrics based on $L_p$ norms have been used in the study of adversarial robustness for $p=0,1,2$, and, $\infty$. It is worth noting that the result of the algorithm depends on the distance metric used. 

Also note that similarity metrics could be used in place of distance metrics. Since they measure how similar two inputs are instead of how different they are, larger values mean smaller distances. However, for simplicity and without loss of generality, we only consider distance metrics in the remainder of this section. 

\begin{list}{Step 3:}
\item \emph{Testing ML model and assessing outputs.} 
\end{list}    
    
The ML model is tested first on the seed test case $t \in T$ and then on the mutants $x \in Mut_t$ sequentially in ascending order of the distance $\|x,t\|$ between the mutant $x$ and the seed $t$; that is, from $x_1$ to $x_n$. 

The process stops when ML model fails on a mutant $x_k \in Mut_t$, i.e. $Fail^M_t(x_k)$. The test case $x_k$ called the \emph{first failure} mutant and denoted by $FF(t)$. The immediate previous mutant $x_{k-1}$ before the first failure mutant $x_k$ is called the \emph{last successful} mutant and denoted by $LS(t)$. See Lines 13 -21 of Alg. \ref{alg:ScenarioDomainAnalysis}. 

Note that it is possible that there are two or more mutants that are of the same distance to the seed $t$. We assume that the sorting in Step 2 puts such mutants in a sequence at random. Thus, the order in which mutants could be tested is not unique. In such cases, there may be more than one mutant that can be found as the first failure (and the last success) in different orderings of mutants. However, the first failure and the last success mutants always have the following properties. 

\begin{lemma}\label{lmm:lmm1}
$\forall x \in Mut_t.\left(Fail^M_t(x) \Rightarrow (FF(t) \preccurlyeq_t x)\right)$. \qed
\end{lemma}

\begin{lemma}\label{lmm:llm2}
$LS(t) \preccurlyeq_t FF(t)$. \qed
\end{lemma}

\begin{lemma}\label{lmm:lmm3}
$\neg \exists x \in Mut_t.\left(LS(t) \prec_t x \prec_t FF(t)\right)$. \qed
\end{lemma}
    
\begin{list}{Step 4:}
\item \emph{Expansion of Mutant Sample Set}
\end{list}

It is possible that the ML model does not fail on any mutants in the set $Mut_t$. In this case, the distances of the mutants in the set $Mut_t$ to $t$ are saved into a variable $listD$ in ascending order. A new set $Mut'_t$ of mutants is generated such that $Mut_t \cap Mut'_t = \emptyset$. The new set $Mut'_t$ of mutants are then saved into variable $mut$ to replace the set $Mut_t$ and the steps (2) - (3) are repeated until a failure mutant is found. See Line 22 - 27 of Alg. \ref{alg:ScenarioDomainAnalysis}. 

After the first failure mutant $f$ is found in $Mut_t$, let $l$ be the last success mutant in $Mut_t$, the following formulas are use to calculate $FF(t)$ and $LS(t)$ for the test case $t$. See Line 30 - 35 of Alg. \ref{alg:ScenarioDomainAnalysis}. 

\begin{eqnarray}
&&FF(t) = \|f, t\|; \label{equ:FFIncr} \\
&&LS(t) = Max\{\|l, t\|, maxOld \}. \label{equ:LSIncr}
\end{eqnarray}
where 
\begin{displaymath}
maxOld = \left\{ 
\begin{array}{ll}
0, &listD = []\\
\underset{x \in listD}{\operatorname{max}}( x \leq \|f, t\|), &otherwise 
\end{array}
\right. 
\end{displaymath}

\begin{list}{Step 5:}
\item \emph{Calculate the Robustness of ML Model $M$}
\end{list}

Finally, the robustness of the ML model is calculated using the following formulas, where the values $LS(t)$ and $FF(t)$ for all $t \in T$ are saved in the variables $listLS$ and $listFF$, respectively; see Line 37 - 38 of Alg. \ref{alg:ScenarioDomainAnalysis}. 

\begin{equation}
R^o= \frac{1}{|T|} \sum_{t \in T}{\|t, FF(t) \|}  
\label{equ:RbyFF}
\end{equation}
\begin{equation}
R^*= \frac{1}{|T|} \sum_{t \in T}{\|t, LS(t) \|} 
\label{equ:RStarbyLS}
\end{equation}

\subsection{Correctness of The Algorithm}

This subsection studies the correctness of the exploration strategy presented in the previous subsection. 

We will first prove that incremental generation and processing of the mutant sample set is equivalent to generating and processing the mutant sample set as a whole, then analyse the conditions under which the algorithm correctly calculates the micro operational robustness. 

Note that the scenario domain surrounding a test case $t$ could be infinite even if it is sparse. Therefore, we have to generate a finite subset of mutants first, and test the ML model to see if the first failure and last success mutants can be found in this subset of mutants. If not, a large subset of mutants is generated. Our algorithm incrementally process the subsets of the mutants by saving the distances of already processed mutants, and finding the first failure $f$ and last success $l$ mutants in the new mutants that have not been processed. After the first failure and last success is found, the first failure $f'$ and last success $l'$ for the whole set of mutants is calculated from the values of $f$ and $l$ and the saved distances of processed mutants. 

Formally, suppose that the mutant sample set is expanded for $K>0$ times in the execution of Alg. \ref{alg:ScenarioDomainAnalysis}, and mutant sets $mut_1, \cdots, mut_{K-1}$ are generated before the first failure mutant $f$ and the corresponding last success mutant $l$ are found in $mut_K$. Then, by Alg. \ref{alg:ScenarioDomainAnalysis}, we have that for all $i \neq j = 1,\cdots, K$,  $mut_i \cap mut_j = \emptyset$ and $\forall x \in \bigcup_{n=1}^{K-1}mut_n.\left(\neg Fail^M_t(x)\right)$. 

An alternative to the above incremental generation and processing the mutants is to generate a large set of mutants that equals to $\bigcup_{n=1}^{K}Mut_n$ at once. Consequently, according to Alg. \ref{alg:ScenarioDomainAnalysis}, the first failure $f'$ and last success $l'$ mutants should be found without expanding the set of mutants. The question is: what is the relationship between $\left< l, f\right>$ and $\left<l',f'\right>$? 

The following theorem gives this relationship so that we can calculate $\left<l',f'\right>$ from $\left< l, f\right>$ using the formula in Equ. (\ref{equ:FFIncr}) and (\ref{equ:LSIncr}). 

\begin{theorem}\label{thm:Expansion}
For any valid orderings of mutants in $mut_K$, there is a valid ordering of mutants in $\bigcup_{n=1}^{K}mut_n$ that preserves the ordering of the mutants in $mut_K$ such that
\begin{enumerate}
\item  $f' = f$.
\item $l' = max\{l, l_o\}$, where 
\[l_o =\underset{x \in mut_{Old}}{\operatorname{argmax}}(x \preccurlyeq f),\] and 
\[mut_{Old} = \bigcup_{n=1}^{K-1}mut_n.\] 
\end{enumerate}
\qed
\end{theorem}

Now, we analyse the correctness of the exploration strategy presented in Alg. \ref{alg:ScenarioDomainAnalysis}. 

First, we formally define the notions of \emph{valid} and \emph{relatively complete} for sets of mutants in the theoretical framework presented in Section \ref{sec:TheoreticalFramework} as follows. 

\begin{definition}\emph{(Validity)}\label{def:validity}

A set $Mut_t$ of mutants of a seed $t$ is \emph{valid}, if every mutant correctly represents the operational scenario $s$ surrounding the seed $t$. Formally, $Mut_t \subset D^s$. \qed
\end{definition}

\begin{definition}\emph{(Completeness)}\label{def:completeness}

A set $Mut_t$ of mutants of seed $t$ is \emph{relatively complete} with respect to a certain distance $\delta_t > 0$, if it contains all points of the sub-domain of the scenario $s$ that surround the seed $t$ within the distance $\delta_t$. Formally, for $\delta_t>0$, 
\[\forall x \in D^s. \|x,t\| \leq \delta_t \Rightarrow (x \in Mut_t).\] 
The maximal distance $\delta_t$ that makes $Mut_t$ relatively complete is called $Mut_t$'s \emph{coverage size} of the scenario $s$ surrounding $t$. \qed
\end{definition}

Theorem \ref{thm:Ro} and \ref{thm:R*} below prove that, under the condition that the coverage size of the mutant set is large enough, $R^o$ and $R^*$ are accurate statistical estimations of the metrics $\rho^{o}$ and $\rho^*$, respectively. 

\begin{theorem}\label{thm:Ro}
If a set $Mut_t$ of $t$'s mutants is valid and relatively complete with a coverage size $\delta_t \geq \|t,FF(t)\|$, then, we have that $\|t,FF(t)\| = \Delta_{D^s}(t,M)$. \qed
\end{theorem} 

Similarly, we have the following theorem about $LS(t)$. 

\begin{theorem}\label{thm:R*}
If a set $Mut_t$ of $t$'s mutants is valid and relatively complete with a coverage size $\delta \geq \|t,FF(t)\|$, then, we have that $LS(t) \preccurlyeq_t \nabla_{D^s}(t,M)$, and there is no $x \in D^s$ such that $LS(t) \prec_t x \prec_t \nabla_{D^s}(t,M)$. \qed
\end{theorem}

Theorem \ref{thm:R*} means that $\|t,LS(t)\|$ is either accurate or the best conservative estimation of $\nabla_{D^s}(t,M)$. Moreover, the theorem states that $\|t,LS(t)\|$ is not accurate (i.e. $\|t,LS(t)\| < \nabla_{D^s}(t,M)$) only if there is $\alpha \in D^s$ such that $\alpha \approx_t FF(t) \wedge  Safe^M_t(\alpha)$. In this case, $\nabla_{D^s}(t,M)=\Delta_{D^s}(t,M)$.

To ensure the accuracy of the results, by Theorem \ref{thm:Ro} and \ref{thm:R*}, we require that the set $Mut_t$ of mutants is valid and complete enough to ensure that its coverage size $\delta_t$ is greater than the minimal distance to failure. 

In Alg. \ref{alg:ScenarioDomainAnalysis}, the mutant sample set is generated incrementally by repeatedly invoking a mutant generation function $GenMuts(t)$ such that 
\begin{eqnarray*}
mut_1 &= &GenMuts^{1}(t), \\
\cdots \\
mut_n &= &GenMuts^{n}(t), \\
\cdots, 
\end{eqnarray*}
where $mut_i \cap mut_j = \emptyset$ for all $i\neq j \geq 0$. $GenMuts^{i}(t)$ denotes calling the mutant generation function $GenMuts$ for $i$ times on a test case $t$. 

Such an incremental mutant set generation function must satisfy the conditions defined below.  

\begin{definition} (Validity and Eventually Completeness of Mutant Generation)\label{def:validCompleteMutantGen}

An incremental mutant generation function $GenMuts$ is \emph{valid}, if for all $t \in D$ and for all $n > 0$, the set $GenMuts^n(t)$ of mutants of $t$ is valid. 

An incremental mutant generation function $GenMuts$ is \emph{eventually relatively complete} with respect to ML model $M$, if, for the minimal integer $K>0$ such that the first failure $f$ is found in the set of generated mutants $GenMuts^K(t)$ but not in $\bigcup_{n=1}^{K-1}GenMuts^n(t)$, the coverage size $\delta_t$ of the set of mutants $\bigcup_{n=1}^{K}GenMuts^n(t)$ is greater than or equals to $\|f,t\|$. 

\qed
\end{definition}

Theorem \ref{thm:Expansion} states that the results of processing mutant set incrementally and calculating $FF(t)$ and $LS(t)$ according to formulas in Equ. (\ref{equ:FFIncr}) and (\ref{equ:LSIncr}) as in Alg. \ref{alg:ScenarioDomainAnalysis} is equivalent to generating and processing the whole mutant set $\bigcup_{n=1}^{K}mut_n$ at once. When the mutant set generation algorithm is valid and eventually complete as in Definition \ref{def:validCompleteMutantGen}, the sets of mutants generated by executing the algorithm Alg. \ref{alg:ScenarioDomainAnalysis}, satisfy the conditions of Theorem \ref{thm:Expansion} as well as Theorem \ref{thm:Ro} and Theorem \ref{thm:R*}. Therefore, we have the following conclusion. 

\noindent\textbf{\textit{Corollary.}}

The exploratory strategy presented in Alg. \ref{alg:ScenarioDomainAnalysis} is correct if the mutant generation function $GenMuts$ used in the algorithm is valid and eventually relatively complete. 
\qed

It is worth noting that, in practice, it may be difficult (if not impossible) to formally prove that a set $Mut_t$ of mutants is valid, because, for most ML applications, the scenario sub-domain $D^s$ is usually not formally defined. Instead, given an algorithm that generates a set $Mut_t$ of mutants from each test case $t$ in the benchmark, its validity has to be validated informally. The validity of the results heavily depends on how mutants are generated. Changing the way that mutants are generated means the scenario is changed, and it will result in different results of the assessment and evaluation of the micro operational robustness, because it represents a different scenario. 

Moreover, the relative completeness of a set of mutants $mut_t$ of a test case $t$ w.r.t. $\Delta_{D^s}(t,M)$, even to find out its coverage size $\delta_t$, is also difficult, if not impossible. Again, the eventual relative completeness of an incremental mutant generation function can only be validated informally. 

\section{An Application to Code Generation}\label{sec:ApplyToCodeGen}

In this section, we apply the proposed method to code generation by giving the technical details of various steps in the proposed method. 

\subsection{The Operational Scenario} 

We are concerned with the uses of LLMs as a tool for code generation. The input to LLMs are NL descriptions of the coding tasks that defines the required functions of the code. The output is expected to be program code written in the required programming language. Here, as a case study, we require the code to be in Java. 

The operational scenario that we are concerned with is, when the user inputs a NL description of a coding task, there are many different words of similar meanings that can be used. The micro operational robustness property of LLMs that we would like to assess is how sensitive LLMs are to such choices of wording in the description of coding tasks. 

In the assessment of the above micro operational robustness property, we limit the scope of the scenario by imposing the following restrictions on the variations of the input to LLMs. In particular, we will not consider any of the following
\begin{itemize}
\item Descriptions of a coding task in sentences of different syntax structures. 
\item Descriptions of a coding task in different languages, such as in German and French. 
\item Descriptions of a coding task using different prompt strategies instead of a basic description of the required functionality. 
\item Descriptions of a coding task that contains spelling errors, which could have been corrected automatically. 
\item Descriptions of a coding task that require the output code in a programming language other than Java. 
\end{itemize} 
Each of the above restrictions could be considered as a different scenario and studied separately. 

However, we do not rule out the situations when a word is not used appropriately in its context, which may cause the NL description to be grammatically and/or semantically incorrect. This is because human users could make such errors and the ability of LLMs to resist them is an important aspect of their robustness in the use of them as coding tools. 

Note that, this overall scenario of code generation can be further divided into many sub-scenarios, for example, according to the topics of coding task, and/or the complexity of the coding task, etc. Such sub-scenarios will be explored in the experiments reported in Section \ref{sec:Experiment}. 

\subsection{Generating Mutants in the Scenario Domain}

From the definition of the scenario above, the input data space for code generation consists of coding tasks described in NL. The mutants are therefore paraphrases of the original descriptions of the coding tasks made by replacing words in the coding task description with words of similar meanings. 

To systematically generate paraphrases of different distances from an original description of the coding task, we employ word embedding techniques, which provide a vector representation of the semantics and usages of words. These techniques have been intensively studied and widely used in NLP; two of the most popular are Word2Vec \citep{mikolov2013} and GloVe \citep{pennington2014}. 

\emph{Word2Vec}, developed by \citet{mikolov2013}, is a ML model that provides a high-dimensional vector for each word in such a way that words with similar meanings have vector representations close to each other in the vector metric space. It also allows for vector arithmetic operations applied to the vector representations of words to reveal their meanings. For example, $\left<king\right> - \left<man\right> + \left<woman\right> = \left<queen\right>$, where $\left<word\right>$ is the vector representation of the word.

\emph{GloVe}, introduced by \citet{pennington2014}, is a model obtained by an unsupervised ML algorithm that generates word embeddings by statistics of word co-occurrences in a text corpus. GloVe constructs a global word-word co-occurrence matrix, capturing the frequency with which words appear together in a given context. GloVe produces dense vector representations of words by factoring the matrix, ensuring that the resulting embeddings reflect both local and global statistical information about the words. 

We utilize GloVe, which is pre-trained, for creating paraphrases rather than Word2Vec because Word2Vec primarily focuses on local context but GloVe captures the relationships between words based on their global co-occurrence in the text. This allows it to facilitate the generation of paraphrases that are both contextually relevant and semantically meaningful. 

The paraphrase generation algorithm is based on the algorithm developed by \citet{zhu2020} to generate high order mutants via combinations of data mutation operators. Let's first introduce some notions and notations. 
		
Let $x$ be a sentence in NL and $w$ be a word in $x$. The operation that replaces an occurrence of the word $w$ in $x$ by another word $w'$ is called the $ReplaceWord$ operation and denoted by $x[w/w']$. This operator is a \emph{data mutation operator} or \emph{datamorphism} \citep{zhu2020}.   
		
A paraphrase $x'= x[w/w']$ obtained by an application of the $ReplaceWord$ operator is a \emph{first order mutant} of $x$. 
		
Let $y$ be a $k$'th order mutant of $x$ obtained by replacing different word occurrences $w_1, \cdots, w_k$ with words $w'_1, \cdots, w'_k$ in $x$. A paraphrase $y'$ obtained by applying the $ReplaceWord$ operator on $y$, i.e. replacing a word $w_{k+1}$ in $y$ by a word $w'_{k+1}$, is an \emph{$(k+1)$'th order mutant}, if $w_{k+1} \neq w_i$ for all $i=1, \cdots, k$. 
		
The paraphrase generation algorithm consists of the following steps. 

\begin{enumerate}
    \item \emph{Sequentialising text}: The original text of the coding task description $x$ in NL is converted into a linear sequence $Seq(x)=\left<w_1, \cdots, w_n \right>$ of words. 
    \item \emph{Retrieving neighbourhood}: Let $n>0$ be a given natural number. For each word $w$ in the sequence $Seq(x)$, the GloVe word embedding model is invoked to produce a set of $n$ words nearest to $w$ in the embedding vector space. Here the set of words is denoted by $Neighbour_n(w)$. 
    \item \emph{Generating paraphrases}: Let $k>0$ be a given natural number and $k\leq L(x)$, where $L(x)$ is the length of the sequence $Seq(x)$. Our paraphrase generation algorithm first generates all first order mutants of the original coding task description by applying the $ReplaceWord$ operator $x[w/w']$, where $w \in Seq(x) \wedge w' \in Neighbour_n(w)$. It then generates all second order mutants by applying the $ReplaceWord$ operator on the first order mutants, and so on, until all $k$'th order mutants are generated. 
\end{enumerate}

For example, consider the coding task description: \textit{Write a Java program to replace a specified character with another character}. Table \ref{tab:Neighbourhood} gives the nearest 4 words to each of the words in the text of the coding task description, where the word with a ranking of 1 is the nearest one according to GloVe. Table \ref{tab:Paraphrases} shows a small sample of paraphrases generated from the sentence with different values of parameter $k$ for the number of words replaced and $n$ for the highest ranking of the replacement words. 

\begin{table*}[t]
\caption{Examples of The Neighbourhoods of Words}\label{tab:Neighbourhood}
\scriptsize
\centering
\begin{tabular}{|c|l|l|l|l|l|l|l|l|l|}
\hline
\textbf{Rank ($n$)} &\textit{write} &\textit{a} &\textit{program} &\textit{to} &\textit{replace} &\textit{specified} &\textit{character} &\textit{with} &\textit{another}\\\hline
1 &writing &another &programs &take &replacing &applicable &characters &and &one\\\hline
2 &read &an &project &would &replacement &parameter &protagonist &while & a\\\hline
3 &publish &one &funding &instead &replaced &specifying &portrayed &made &an\\\hline
4 &notes &same &programme &could &replaces &depending &portrays &both &same\\\hline
\end{tabular} 
\end{table*}

\begin{table*}
\caption{Examples of Paraphrases}\label{tab:Paraphrases}
\centering
\scriptsize
\begin{tabular}{|l|c|c|}
\multicolumn{3}{l}{\textbf{Original}: \textit{Write a Java program to replace a specified character with another character.}} \\\hline
\textbf{Paraphrase} &$k$ &$n$\\\hline
\textit{Writing} a Java program to replace a specified character with another character.&1 &1\\\hline
Write a Java program to replace a \textit{applicable} character with another character. &1 &1 \\\hline
Write a Java program to replace a specified character with \textit{one} character. &1 &1\\\hline
Write a Java program to replace a specified character with \textit{a} character.&1 &2\\\hline
Write a Java program to replace a \textit{parameter} character with another character. &1 &2\\\hline
Write a Java program to replace \textit{one} specified character with another character. &1 &3\\\hline
Write a Java program to replace a specified \textit{protagonist} \textit{while} another character.&2 &2\\\hline
Write a Java program to replace \textit{an} \textit{applicable} character with another character. &2 &2 \\\hline
Write a Java program to replace a \textit{parameters} character with \textit{a} character.&2 &2\\\hline
Write \textit{an} Java program \textit{would} replace a \textit{parameter} character with another character.&3 &2\\\hline
\textit{Notes} a Java program to replace \textit{same} \textit{depending} character with another character. &3 &4\\\hline
\textit{Publish} a Java program to replace \textit{one} specified \textit{portrayed} \textit{made} another character. &4 &3\\
\hline 
\end{tabular} 
\end{table*}

As shown in this example, the semantic similarity between a paraphrase and the original sentence cannot be easily determined by the parameters $n$ and $k$. Thus, it is important to order the set of paraphrases by measuring their similarity or distance to the original coding task description. 

\subsection{Measuring Distances from Mutants to the Seed}

At this step, each paraphrase generated in the previous step is measured for its distance to the original task description using a similarity metric between NL sentences. 

In linguistics, the similarity between sentences can be measured not only based on the geometric distance, but also based on the lexical overlapping, structure and semantics as well. 

Text similarity metrics can be categorized into the following types. 
\begin{itemize}
\item \emph{Overlap-based metrics} measure similarity by counting common elements such as words or n-grams between texts and measuring the extent of n-gram overlap. Examples in this category include the metrics \emph{BLEU} \citep{papineni2002}, \emph{ROUGE} \citep{lin2004}, \emph{Meteor} \citep{banerjee2005}, and \emph{ChrF} \citep{popovic2015}.
\item \emph{Distance-based metrics}  embed texts in a vector space and quantify the similarity between texts by measuring either the spatial or angular distance between these vector representations in the embedding vector space. A typical example is the $L_2$ norm, which is most commonly used because it measures the \emph{Euclidean} distance and captures both the magnitude and direction of differences between vectors. Another example is the cosine similarity metric, which calculates the \emph{Cosine} of the angle of two vectors that represent the embeddings of the sentences in comparison. 
\item \emph{Edit distance metrics} quantify the number of editing operations required to transform one text into another. These operations typically include insertions, deletions, and substitutions. \emph{Levenshtein} distance is a widely-used edit distance metric, and it is classified into word-level and character-level types. The character-level distance calculates the minimum number of single-character edits needed to convert one text to another, while the word-level distance measures edits at the word level. 
\item \emph{Semantics-based metrics} employ ML models that map a pair of sentences to a similarity score according to the similarity in their meanings. For example, \emph{STSb-RoBERTa-base}\footnote{https://huggingface.co/cross-encoder/stsb-roberta-base} is such a ML model in the cross-encoder architecture RoBERTa \citep{liu2019roberta} fine-tuned on the Semantic Textual Similarity (STS) benchmark dataset to produce semantic similarity scores for pairs of text. In this paper, it is referred to as the \emph{RoBERTa} metric.

\end{itemize}

The metrics used and studied in this paper are: \emph{Euclidean}, \emph{Cosine}, \emph{BLEU}, \emph{ROUGE}, \emph{Meteor}, \emph{ChrF}, \emph{Levenshtein}, and \emph{RoBERTa}. 

\subsection{Testing LLMs and Assessing Generated Code}

In this step, the ML model is tested first on the original coding task description $t$, and then on its paraphrases sequentially one by one in ascending order of distance from $t$. For each paraphrase $x$ of $t$, the algorithm queries the LLM under test with the paraphrase $x$ as the input, and checks if the code generated by the model is functionally equivalent to the code generated from the original task description $t$. 

The process terminates when it finds a paraphrase for which the generated code is not functionally equivalent to the original code. This paraphrase is referred to as the first failed paraphrase and denoted by $FF(t)$, and the paraphrase just before the first failed one is called the last successful paraphrase and denoted by $LS(t)$. These are used to calculate the $R^o$ and $R^*$ metrics, respectively, later in Step 4. This process is repeated on all code generation tasks in the test dataset. 

It is worth noting that the functional equivalence between two programs is not computable in general. We take a pragmatic approach by automatically generating test data from both programs and test the programs on the union of the test data; see \citep{ghosh2024} for more details. Formally, the predicate $Fails^M_t(x)$ is defined as follows. 

\[Fails^M_t(x) \Leftrightarrow \exists d \in TS(P) \cup TS(P'). (P(d) \neq P'(d)). \]
Here, $P=M(t)$ and $P'=M(x)$ are programs generated by ML model $M$, and $TS(P)$ and $TS(P')$ are the sets of functional test cases generated for $P$ and $P'$, respectively. 

When the testing of the ML model $M$ finishes, we have the data $\{ \left<LS(t),FF(t)\right> ~|~ t \in T \}$. Then, the metrics $R^{o}$ and $R^*$ for model $M$ are calculated using the formulas (\ref{equ:RbyFF}) and (\ref{equ:RStarbyLS}), respectively. 

The algorithm Alg. \ref{alg:ScenarioDomainAnalysis} with the instantiation of the key components presented above have been implemented as a datamorphic test system written in Java in the Morphy test automation environment \cite{zhu2019a,zhu2019,zhu2020}. Details of the implementation are omitted in this paper for the sake of space, but can be found in \cite{PaulPhDThesis}. 

\section{Experiments}\label{sec:Experiment}

This section reports the experiments with the proposed method to answer the following research questions. 

\begin{description}
\item [RQ1] (\emph{Effectiveness}): Is the proposed method capable of providing meaningful results that distinguish LLMs according to their robustness in code generation?
\item [RQ2] (\emph{Efficiency}): Is the proposed method efficient enough for practical use?
\end{description}

\subsection{Design of The Experiments}\label{sec:designOfExperiment}

Four LLMs were chosen as the subject in the experiment for their availability and low cost of access as well as their wide usage in practice for code generation at the time when the experiments started. The LLMs were \emph{Llama2-13B}, \emph{Codex}, \emph{Gemini Pro 1.0}, and \emph{Falcon}. 

\begin{figure}[ht]
\centering
\includegraphics[width=7cm]{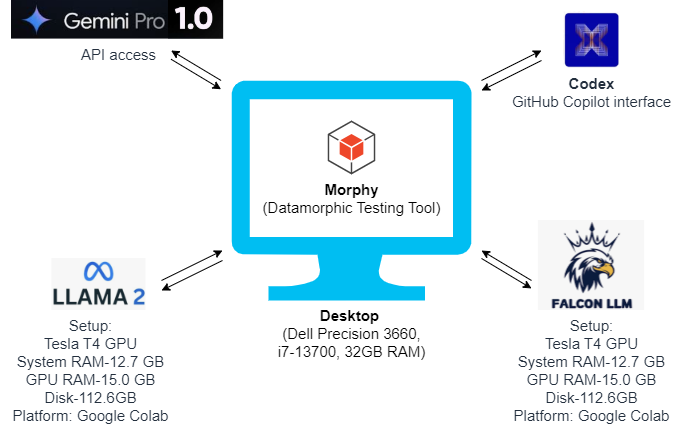}  
\caption{The Experiment Platform}
\label{fig:fig4_1}
\end{figure}

The experiments were conducted with automated ML testing tool Morphy \cite{zhu2019,zhu2020} running on a desktop computer accessing LLM models running on the cloud through API invocations. Specifically, GeminiPro was executed via its API provided by Google. Falcon and Llama2 were deployed on Google Colab with GPU support, and Codex was accessed through GitHub Copilot. Figure \ref{fig:fig4_1} illustrates the configuration of the experiment platform.

We formed a test dataset $D_1$ by selecting 900 coding tasks at random from the ScenEval benchmark \citep{ghosh2024}. 
Table \ref{tab:StatisticsOfTestDataset} gives the statistical characteristics of the test dataset $D_1$. 

\begin{table}[htbp]
\centering
\caption{Statistical Characteristics of the Test Dataset $D_1$}
\label{tab:StatisticsOfTestDataset}
\begin{scriptsize}
\begin{tabular}{|l|c|c|}
\hline
\textbf{Feature} & \textbf{\#Textbook Tasks} & \textbf{\#Real Tasks}\\ \hline\hline
Number of Topics Covered & 9 &7\\ \hline
Average Num of Tasks in A Topic & 66.67 & 42.85\\ \hline
Max Num of Tasks in A Topic & 156 & 61\\ \hline
Min Num of Tasks in A Topic & 8 & 5\\ \hline\hline
Average Complexity & 3.448 &3.200\\ \hline
Max Complexity & 13 &6\\ \hline
Min Complexity & 1 &1\\ \hline\hline
Average Length of Input & 16.800 &25.344 \\\hline
Max Length of Input & 35 &31 \\ \hline
Min Length of Input & 11 & 18\\ \hline\hline
Total Number of Tasks & 600 &300\\ \hline\hline
\end{tabular}
\end{scriptsize}
\end{table}

The testing was conducted in the process described in Section \ref{sec:ProposedMethod} for each LLM and repeated for each of the similarity metrics used to measure the distance between the paraphrases and the original coding task description. The experiment data and their analysis are reported in the following subsections to answer the research questions. 

\subsection{Effectiveness}

To answer research question \emph{RQ1}, we first apply the method on the whole test dataset to evaluate and compare the robustness of the subject LLMs to demonstrate that the proposed method can distinguish LLMs on their robustness of code generation. Then, we apply the method again to various scenarios to demonstrate that LLMs can be distinguished on their robustness in different scenarios. We also provide a comprehensible interpretation of the results. 

\subsubsection{Comparing LLMs on Overall Robustness} 

To compare the robustness of the LLMs, we apply the $R^o$ and $R^*$ metrics on the whole dataset of the experiment. Table \ref{tab:combined_robustness} gives the experiment results including the average values of $R^o$ and $R^*$ for each LLM and distance metric over the whole set of test data with the margins of errors (MoE) at the confidence level equals to $95\%$. Details of the statistical analysis of the margins of errors can be found in Appendix C. 

\begin{table*}[ht]
\centering
\caption{Robustness Measured by $R^o$ and $R^*$ with Different Similarity Metrics on Paraphrases}\label{tab:combined_robustness}
\resizebox{\textwidth}{!}{
\setlength{\tabcolsep}{1pt}
\begin{scriptsize}
\begin{tabular}{|l|c|c|c|c|c|c|c|c|} 
\hline
\multirow{2}{*}{\textbf{Similarity Metric}} 
& \multicolumn{2}{c|}{\textbf{Gemini-Pro}} 
& \multicolumn{2}{c|}{\textbf{Codex}} 
& \multicolumn{2}{c|}{\textbf{Llama2}} 
& \multicolumn{2}{c|}{\textbf{Falcon}} \\ \cline{2-9} 
& $R^o \pm \text{MoE}$& $R^* \pm \text{MoE}$& $R^o \pm \text{MoE}$& $R^* \pm \text{MoE}$& $R^o \pm \text{MoE}$ &$R^* \pm \text{MoE}$ & $R^o \pm \text{MoE}$ & $R^* \pm \text{MoE}$\\ \hline
RoBERTa & $1.0365 \pm 0.0202$ & $1.1117 \pm 0.0189$ & $1.0879 \pm 0.0194$ & $1.1432 \pm 0.0205$ & $1.1945 \pm 0.0191$ & $1.2478 \pm 0.0204$ & $1.2046 \pm 0.0197$ & $1.2550 \pm 0.0209$ \\ \hline
Euclidean & $0.5307 \pm 0.0120$ & $0.4991 \pm 0.0191$ & $0.5103 \pm 0.0131$ & $0.4814 \pm 0.0128$ & $0.4021 \pm 0.0110$ & $0.3748 \pm 0.0107$ & $0.4065 \pm 0.0091$ & $0.3781 \pm 0.0097$ \\ \hline
Cosine Similarity & $0.9291 \pm 0.0025$ & $0.9302 \pm 0.0024$ & $0.9301 \pm 0.0027$ & $0.9381 \pm 0.0024$ & $0.9801 \pm 0.0020$ & $0.9811 \pm 0.0019$ & $0.9703 \pm 0.0022$ & $0.9712 \pm 0.0020$ \\ \hline
BLEU & $0.4273 \pm 0.0125$ & $0.4565 \pm 0.0141$ & $0.4164 \pm 0.0105$ & $0.4388 \pm 0.0119$ & $0.5336 \pm 0.0125$ & $0.5628 \pm 0.0148$ & $0.5295 \pm 0.0124$ & $0.5583 \pm 0.0149$ \\ \hline
ROUGE-N & $0.6224 \pm 0.0103$ & $0.6428 \pm 0.0101$ & $0.6354 \pm 0.0120$ & $0.6587 \pm 0.0112$ & $0.7414 \pm 0.0125$ & $0.7662 \pm 0.0122$ & $0.7381 \pm 0.0114$ & $0.7606 \pm 0.0111$ \\ \hline
ROUGE-L & $0.6950 \pm 0.0091$ & $0.7130 \pm 0.0088$ & $0.7066 \pm 0.0089$ & $0.7235 \pm 0.0080$ & $0.8104 \pm 0.0079$ & $0.8277 \pm 0.0065$ & $0.8085 \pm 0.0058$ & $0.8242 \pm 0.0055$ \\ \hline
METEOR & $0.6552 \pm 0.0106$ & $0.6785 \pm 0.0107$ & $0.6692 \pm 0.0112$ & $0.6925 \pm 0.0105$ & $0.7753 \pm 0.0109$ & $0.7995 \pm 0.0105$ & $0.7718 \pm 0.0107$ & $0.7939 \pm 0.0103$ \\ \hline
ChrF & $66.9836 \pm 0.7078$ & $68.8565 \pm 0.7017$ & $68.0947 \pm 0.7910$ & $69.8502 \pm 0.7738$ & $68.6146 \pm 0.7879$ & $70.2630 \pm 0.7705$ & $68.3506 \pm 0.7827$ & $70.0672 \pm 0.7712$ \\ \hline
Levenshtein-Word & $1.1343 \pm 0.0253$ & $1.0233 \pm 0.0210$ & $1.1288 \pm 0.0209$ & $1.0151 \pm 0.0168$ & $1.1260 \pm 0.0209$ & $1.0096 \pm 0.0168$ & $1.1248 \pm 0.0184$ & $1.0551 \pm 0.0150$ \\ \hline
Levenshtein-Char & $11.5151 \pm 0.4090$ & $10.1133 \pm 0.4933$ & $10.5876 \pm 0.4901$ & $9.4900 \pm 0.4834$ & $10.1340 \pm 0.4701$ & $9.1700 \pm 0.4842$ & $10.3320 \pm 0.4835$ & $9.3000 \pm 0.4798$ \\ \hline
\end{tabular}
\end{scriptsize}
}
\end{table*}

\begin{figure}[htb]
\centering
\includegraphics[width=8cm]{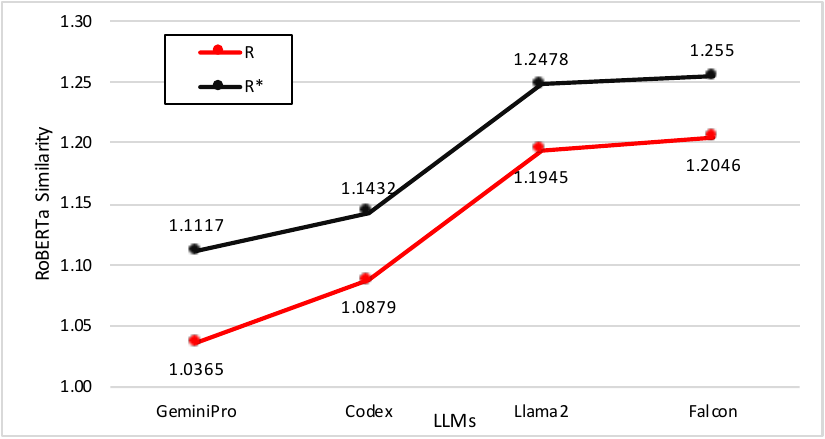} \\ 
\scriptsize{(a) With RoBERTa Similarity} \\
\includegraphics[width=8cm]{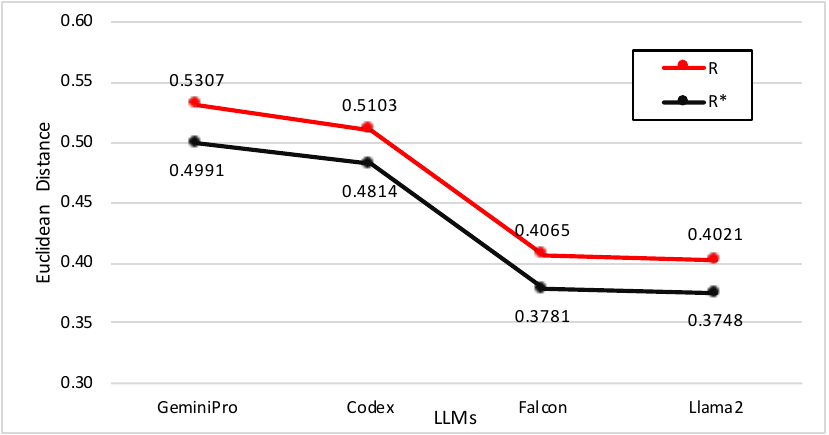} \\ 
\scriptsize{(b) With Euclidean Distance}
\caption{Comparison of LLMs on Robustness of Code Generation}
\label{fig3}
\end{figure}

Figure \ref{fig3}(a) and (b) depict the robustness of the LLMs under two similarity metrics RoBERTa and Euclidean, respectively, where the red lines are the average $R^{o}$ scores calculated from the first failed solutions and the black lines are the average $R^*$ scores calculated from the last successful solutions. 

Note that, when using Euclidean distance between two texts as a measure of similarity, a lower score means that the two texts are more similar to each other so a smaller  $R^{o}$ ($R^*$) value means a poorer robustness. In contrast, when using RoBERTa to measure similarity, a higher score means the text are more similar to each other so poorer robustness is indicated by a \textit{higher} $R^o$ or $R^*$ value because it means that the code generation model fails on more similar inputs. 

\begin{observation} 
The proposed domain analysis method for robustness testing and evaluation is capable of providing accurate measurements of the robustness of LLMs in code generation with all of the similarity metrics used. 
\end{observation} 

Another observation based on the data presented in Table \ref{tab:combined_robustness} is that the rankings of LLMs according to robustness metrics $R^o$ and $R^*$ are consistent across all similarity metrics, although the specific values of these metrics on each LLM vary with the similarity metrics used. In particular, the data suggests that Gemini-Pro is the best on robustness as it scores highest on $R^o$ with all similarity metrics. Codex is second best as it has the highest scores on only one similarity metric (BLEU), but scored second on all others. 

These observations were validated through paired two-tailed hypothesis $t$-tests on whether a LLM $A$ is different from $B$ on robustness according to $R^o$. The results on $R^o$ with RoberTa showed that the validity of following statements were statistically significant with the confidence level equals to 95\% and the $p$-value less than 0.05. 
\[ Gemini Pro \succ Codex, Gemini Pro \succ Llama 2 \]
\[ Gemini Pro \succ Falcon, Codex \succ Falcon \]
\[ Codex \approx Llama 2, Falocn \approx Llama 2\]
where $A \succ B$ means LLM $A$ is significantly stronger on robustness according to $R^o$, while $A \approx B$ means that the different between LLMs $A$ and $B$ on robustness according to $R^o$ is not statistically significant. Details of the hypothesis tests can be found in Appendix B. 

The hypothesis tests also revealed that all other distance and similarity metrics were consistent with RoberTa on statements $Codex \approx Llama 2$, $Falocn \approx Llama 2$, and $Codex \succ Falcon$. However, none of them can distinguish Gemini Pro from other LLMs as Roberta does, except that Levenshtein-Char can validate $Gemini Pro \succ Llama 2$. Moreover, distance metrics RougeL and Levenshtein-Word were unable to distinguish any of the LLMs. This implies that the underlying similarity metrics are not equally capable of distinguishing LLMs. Therefore, the following conclusion can be drawn. 

\begin{observation} 
The similarity metrics used in $R^o$ ($R^*$) metric have different capability of distinguishing LLMs on robustness. RoberTa is the most powerful one among the metrics studied in the experiment. 
\end{observation} 

\subsubsection{Comparing Robustness in Various Sub-Scenarios}

In order to apply the proposed method to various sub-scenarios in which LLMs are employed for code generation, the dataset $D_1$ is split into a number of subsets such that each subset contains the coding tasks of the sub-scenario. 
The benchmarks for these sub-scenarios are formed by grouping the coding task descriptions according to the complexity and topic of the coding task. 

The complexity of a coding task is measured by the cyclomatic complexity of the reference solution provided by the ScenEval benchmark. A total of 6 subsets were formed with complexity ranging from 1 to 6. Then the model's robustness was evaluated using $R^o$ and $R^*$ metrics on each subset the same way as for the whole dataset. 

\begin{figure}[ht]
\centering
\includegraphics[width=8cm]{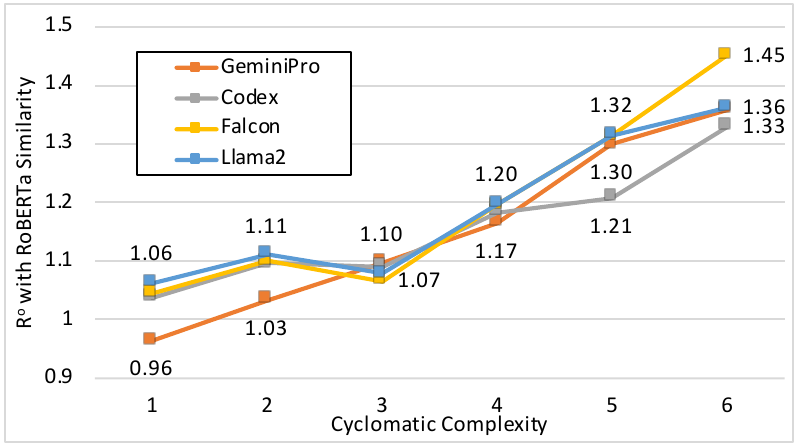} \\ 
 \scriptsize{(a) RoBERTa Similarity} \\
\includegraphics[width=8cm]{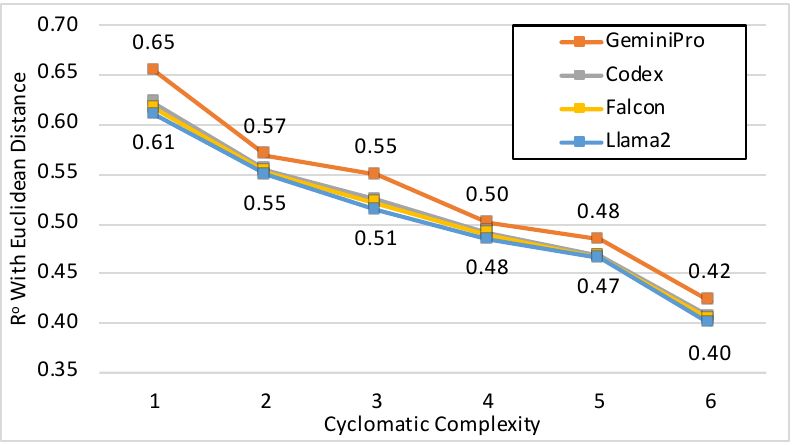} \\ 
\scriptsize{(b) Euclidean Distance} \\
\caption{Variation of Robustness with Cyclomatic Complexity}
\label{fig4}
\end{figure} 

Figure \ref{fig4}(a) and (b) show how robustness varies with the increase of cyclomatic complexity when the similarity is measured by RoBERTa and Euclidean, respectively. In both cases, our experiment data clearly suggested that the robustness decreases when complexity increases as one could expect. This phenomenon is also clearly observed for all other similarity metrics as well. In fact, the Pearson correlation coefficients between cyclomatic complexity and the robustness score $R^o$ are highly correlated; see Table \ref{tab:CorrelationComplexity} for the Pearson correlation coefficients between cyclomatic complexity and robustness $R^o$ with various distance metrics. 

\begin{table}[htbp]
\centering
\caption{Pearson Correlation Coefficients btw Cyclomatic Complexity and Robustness Measure $R^o$}
\label{tab:CorrelationComplexity}
\begin{scriptsize}
\begin{tabular}{|l|c|c|c|c|}
\hline
\textbf{Similarity Metrics} &\textbf{Codex} &\textbf{GeminiPro} &\textbf{Falcon}&\textbf{Llama2}\\ \hline
RoBERTa&0.9924&0.9594&0.9368&0.9404\\ \hline
BLEU&0.8577&0.8860&0.8989&0.9007\\ \hline
ChrF&0.8287&0.7063&0.8950&0.8796\\ \hline
Cosine&0.7661&0.7610&0.7198&0.7706\\ \hline
Meteor&0.8171&0.8057&0.8188&0.8116\\ \hline
RougeL&0.8483&0.8838&0.8477&0.8498\\ \hline
RougeN&0.8243&0.8699&0.7700&0.7962\\ \hline
Euclidean&-0.9795&-0.9869&-0.9873&-0.9849\\ \hline
Levenshtein-Char&-0.7751&-0.8100&-0.8322&-0.8343\\ \hline
Levenshtein-Word&-0.7252&-0.7158&-0.9029&-0.8960\\ \hline
\end{tabular}
\end{scriptsize}
\end{table}

\begin{observation}
The robustness of a LLM in code generation decreases when the complexity of the coding task increases. 
\end{observation}

\begin{observation}
The proposed method can distinguish the robustness of LLMs in code generation on different scenarios defined according to coding task complexity. 
\end{observation}

To evaluate the robustness of code generation in scenarios of different coding topics, six subsets of $D_1$ were constructed by filtering the topic of the coding task on six distinct topics: Array, String, OOP, Regular Expression, Multi-threading, and Data Structure (DS). This enabled us to compare the model's robustness across different topics. 

\begin{figure}[ht]
\centering
\includegraphics[width=8cm]{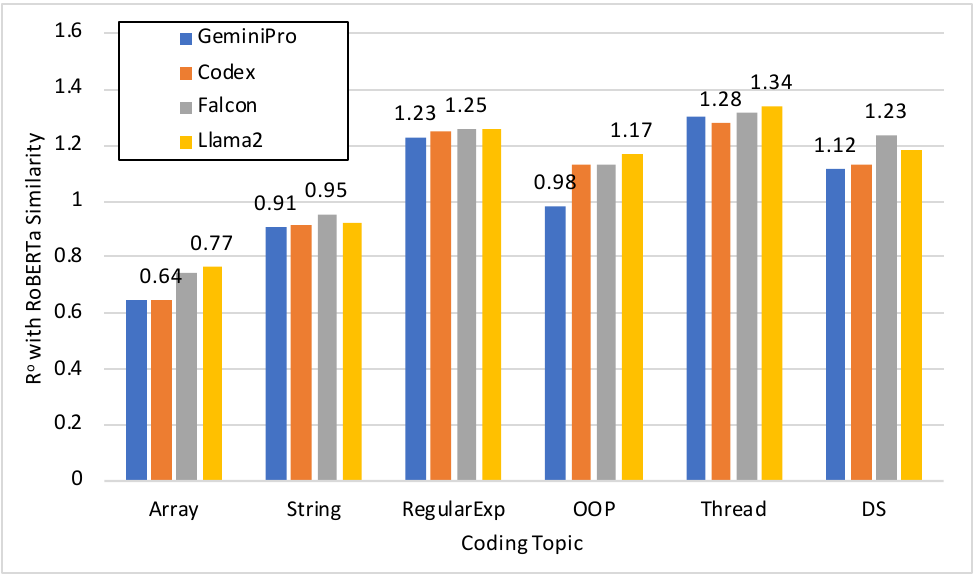}  \\ 
\scriptsize{(a) RoBERTa Similarity} \\
\includegraphics[width=8cm]{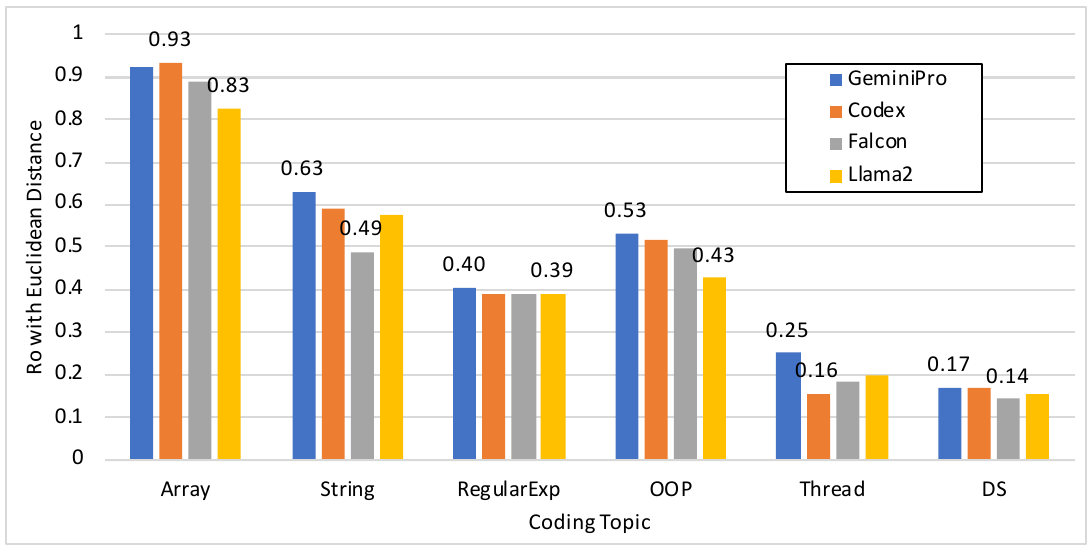} \\ 
\scriptsize{(b) Euclidean Distance} \\
\caption{Variation of Robustness with Topics}
\label{fig5}
\end{figure}

Figure \ref{fig5} (a) and (b) show how robustness varies with the topics when the similarity is measured by RoBERTa and Euclidean, respectively. It is observed that robustness scores vary across topics for various models. In particular, as we expected, the robustness scores demonstrated that the model robustness are poorer on the advanced coding tasks like multi-threading, regular expressions, and complex problems such as data structures rather than simpler coding tasks such as array and string processing tasks.

From the experimental data, we can make the following observations. 

\begin{observation} 
The proposed method is capable of distinguishing ML models in their robustness on various scenarios defined by coding topic. 
\end{observation}

\begin{observation}
LLMs performed better in robustness on basic and simple coding tasks than advanced and complicated coding tasks. 
\end{observation}

\subsubsection{Interpretation of Robustness Metrics}

To understand the meanings of the robustness metrics, we analyse the tipping points obtained in the process of testing and evaluating robustness. 

In general, a pair consisting of the last success mutant and the first failure mutant is called a \emph{tipping point}. This is where a model switches from success to failure as the ML model moves away from the original test case. In case of code generation, it means that the ML model switches from generating a correct program to generating an incorrect program even though the difference between the input coding task descriptions are very small. 

We now analyse the difference between the programs $LS$ and $FF$ generated by LLMs on such pairs of mutants to answer two research questions: (a) are the programs generated by an LLM before and after a tipping point similar to each other syntactically and semantically? (b) when will a tipping point happen? 

For question (a), we use code editing distance to measure the similarity between two pieces of code by counting the minimum number of editing operations required to transform one piece of code into the other. 

In particular, we use the editing distances based on Abstract Syntax Trees (ASTs) to reflect the syntax distances, and data flow graphs (DFG) and Program Dependency Graphs (PDGs) for more semantics-oriented distances. 

Let $t \in T$ be a seed test case and $m$ be AST, DFG or PDG. We write $Dist^{LS}_{m}(t)$ to denote the distance between the reference solution of the seed test case $t$ and the program generated from the last successful mutant of the seed $t$. Similarly, we write $Dist^{FF}_{m}(t)$ for the distance between the reference solution of the seed test case $t$ and the program generated from the first failed mutant. 
We then calculate the differences between $Dist^{LS}_{m}(t)$ and  $Dist^{FF}_{m}(t)$ as follows. 
\begin{equation}
Diff_m(t) = Dist^{FF}_{m}(t) - Dist^{LS}_{m}(t)
\end{equation}

\begin{figure*}[htb]
\centering
\includegraphics[width=12cm]{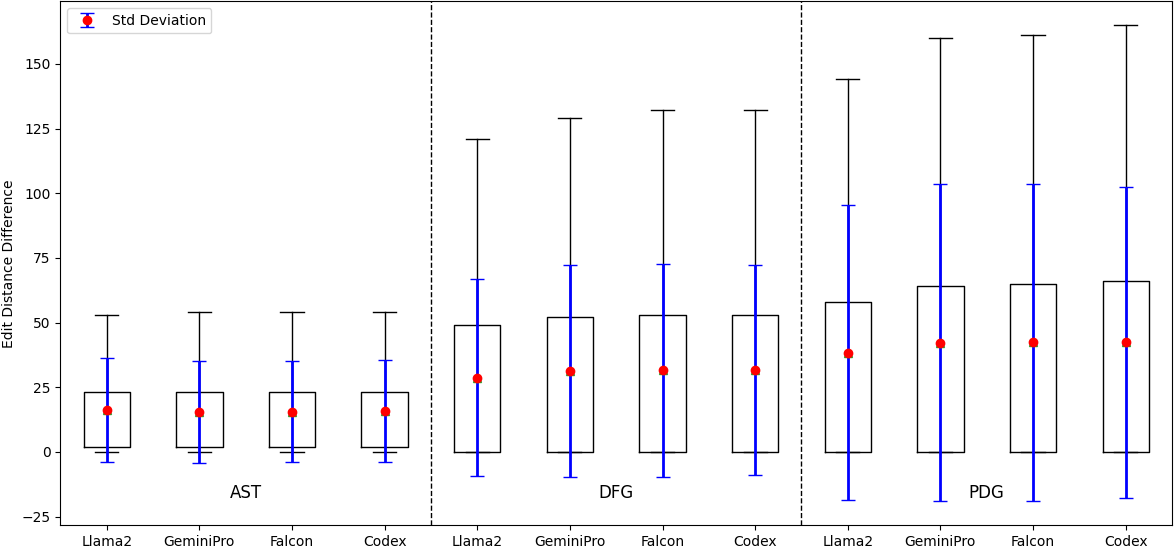} \\ 
\caption{Differences btw the Edit Distances to the Original Description before and after the Tipping Points}
\label{fig6}
\end{figure*}

The statistical results over all test cases in $T$ are shown in Figure \ref{fig6} with the maximum (the upper black bar), minimum (the lower black bar), average (the red dot), and standard deviation (the blue line segment), and the central quartiles (the black box) for three different code editing distance metrics AST, DFG and PDG.  

The data demonstrate that, for all LLMs, the average values of $Diff(t)$ are positive for all LLMs and well above zero. And, most of the data are distributed above the zero. Thus, we have the following observations. 

\begin{observation}
On average, the last successful solution generated by a LLM is closer to the reference solution than the first failed solution when the paraphrase increases its distance to the original coding task description. 
\end{observation}

The distribution of the differences in the distances to the reference solutions before and after tipping points as shown in Figure \ref{fig6} shows that on average there is a big jump in the distances to the reference solution before and after the tipping points. 

\begin{observation}
In most cases, there is a big increase in the distance to the reference solution before and after a tipping point.
\end{observation}

Although the above two observations hold on average, there are exceptional cases and we have made the following observation. 

\begin{observation}
The distance to reference solution may decrease after a tipping point. In other words, it is possible that the first failure solution has a distance closer to the reference solution than the last success solution. 
\end{observation} 

The second analysis of the tipping points aims to provided a comprehensible explanation of the $R^{o}$ and $R^*$ metrics. 
We analysed the paraphrases of the first failed mutants to find out how many words were changed and the semantic similarity between the original words and the replacement words. 

In order to achieve this goal, we calculated the following two values of the paraphrases from each first failed mutant. 
\begin{itemize}
\item $k$: the number of words in the original text description of coding task in the seed test case replaced by different words to generate the paraphrase of the mutant test case. 
\item $n$: the largest distance between a word in the original text description of coding task of the seed and the word used to replace it to generate the paraphrase; here, $n$ is the order of the replacement word in the neighbourhood of the original word in the word embedding vector space.  
\end{itemize}

Table \ref{tab:averageNandK} shows the average values of $n$ and $k$ over all test cases for different metrics used in the robustness evaluation. 

\begin{table*}[ht]
\centering
\caption{Average Values of $n$ and $k$ of the First Failure Paraphrases}
\label{tab:averageNandK}
\resizebox{\textwidth}{!}{ 
\begin{scriptsize}
\begin{tabular}{|l|c|c|c|c|c|c|c|c|}
\hline
\multirow{2}{*}{\textbf{Metric}} &\multicolumn{2}{c|}{\textbf{Codex}} &\multicolumn{2}{c|}{\textbf{GeminiPro} }&\multicolumn{2}{c|}{\textbf{Llama2}}&\multicolumn{2}{c|}{\textbf{Falcon}}\\ \cline{2-9}
 & $n \pm \text{MoE}$ & $k \pm \text{MoE}$ & $n \pm \text{MoE}$ & $k \pm \text{MoE}$ & $n \pm \text{MoE}$ & $k \pm \text{MoE}$ & $n \pm \text{MoE}$ & $k \pm \text{MoE}$ \\ \hline
RoBERTa & $3.67 \pm 0.14$ & $2.03 \pm 0.04$ & $3.86 \pm 0.16$ & $2.73 \pm 0.05$ & $2.65 \pm 0.12$ & $2.43 \pm 0.04$ & $2.67 \pm 0.13$ & $2.03 \pm 0.06$ \\ \hline
ChrF & $2.67 \pm 0.14$ & $2.62 \pm 0.05$ & $2.68 \pm 0.15$ & $2.03 \pm 0.04$ & $2.46 \pm 0.11$ & $2.11 \pm 0.03$ & $2.64 \pm 0.12$ & $2.22 \pm 0.04$ \\ \hline
Euclidean & $2.76 \pm 0.15$ & $2.73 \pm 0.05$ & $2.67 \pm 0.14$ & $2.48 \pm 0.04$ & $2.47 \pm 0.12$ & $2.32 \pm 0.04$ & $2.68 \pm 0.13$ & $2.03 \pm 0.04$ \\ \hline
Cosine & $2.86 \pm 0.16$ & $2.23 \pm 0.04$ & $2.67 \pm 0.14$ & $2.33 \pm 0.04$ & $2.48 \pm 0.12$ & $2.41 \pm 0.04$ & $2.38 \pm 0.11$ & $2.33 \pm 0.04$ \\ \hline
BLEU & $2.63 \pm 0.14$ & $2.09 \pm 0.04$ & $2.22 \pm 0.12$ & $2.63 \pm 0.05$ & $2.63 \pm 0.13$ & $2.03 \pm 0.03$ & $2.63 \pm 0.13$ & $2.04 \pm 0.04$ \\ \hline
METEOR & $2.68 \pm 0.15$ & $2.03 \pm 0.04$ & $2.45 \pm 0.13$ & $2.43 \pm 0.04$ & $2.58 \pm 0.12$ & $2.55 \pm 0.04$ & $2.62 \pm 0.13$ & $2.38 \pm 0.04$ \\ \hline
Levenshtein-Char & $2.57 \pm 0.15$ & $2.03 \pm 0.04$ & $2.56 \pm 0.12$ & $2.03 \pm 0.04$ & $2.57 \pm 0.13$ & $2.03 \pm 0.03$ & $2.57 \pm 0.13$ & $2.03 \pm 0.04$ \\ \hline
Levenshtein-Word & $3.60 \pm 0.18$ & $1.42 \pm 0.03$ & $3.56 \pm 0.17$ & $1.33 \pm 0.03$ & $2.46 \pm 0.11$ & $1.09 \pm 0.02$ & $2.42 \pm 0.12$ & $1.14 \pm 0.02$ \\ \hline
ROUGE-N & $2.77 \pm 0.15$ & $2.20 \pm 0.04$ & $2.99 \pm 0.16$ & $2.16 \pm 0.04$ & $2.46 \pm 0.11$ & $2.11 \pm 0.03$ & $2.33 \pm 0.11$ & $2.12 \pm 0.04$ \\ \hline
ROUGE-L & $2.57 \pm 0.13$ & $2.32 \pm 0.04$ & $2.77 \pm 0.15$ & $2.34 \pm 0.04$ & $2.41 \pm 0.11$ & $2.14 \pm 0.03$ & $2.21 \pm 0.10$ & $2.03 \pm 0.03$ \\ \hline
\textbf{Average} & $\mathbf{2.88 \pm 0.15}$ & $\mathbf{2.17 \pm 0.04}$ & $\mathbf{2.84 \pm 0.15}$ & $\mathbf{2.25 \pm 0.04}$ & $\mathbf{2.52 \pm 0.12}$ & $\mathbf{2.12 \pm 0.03}$ & $\mathbf{2.52 \pm 0.12}$ & $\mathbf{2.04 \pm 0.03}$ \\ \hline
\end{tabular}
\end{scriptsize}
}
\end{table*}

From the data presented in Table \ref{tab:averageNandK}, we have the following observation. 

\begin{observation}
Assume that a LLM generates a correct program from a NL description of the coding task. Then, on average, the LLM will generate an incorrect program code if one changes 2 to 3 words in the description of the coding task by replacing them with words that are the 2nd to 3rd closest to the original words in meaning. If one changes the description smaller than that, the LLM will most probably still be able to generate correct code. 
\end{observation}

In conclusion, from the above observations, we can conclude that the proposed method is effective in the sense that it is capable of providing meaningful measurements of LLMs that can distinguish different LLMs on robustness in code generation and the robustness of a LLM on various coding scenarios. 

\subsection{Efficiency}

We now address research question \emph{RQ2} on the efficiency of the proposed method. We will assess the feasibility of the proposed method by the computational resource required in terms of computation time, memory space, and the cost of using the LLMs under test. 

\subsubsection{Cost of Using LLMs} 

First, we measure the cost of using the LLM under test using the number of queries to the ML models for each similarity metric. The data on the average numbers of LLM queries for each of the four LLMs and various distance metrics are given in Table \ref{tab:AverageLLMQueries} together with the margins of errors at the confidence level $95\%$. 

\begin{table}[ht]
\centering
\caption{Average Number of LLM Queries Per Test Case}
\label{tab:AverageLLMQueries}
\resizebox{\columnwidth}{!}{
\begin{scriptsize}
\begin{tabular}{|l|c|c|c|c|c|c|c|}
\hline
\textbf{Metric} & \textbf{GeminiPro} &\textbf{Codex} &\textbf{Falcon} &\textbf{Llama2}\\ \hline
BLEU             & $14.82 \pm 1.85$ & $14.13 \pm 1.82$ & $13.79 \pm 1.78$ & $13.65 \pm 1.75$ \\ \hline
ChrF             & $16.02 \pm 1.94$ & $15.34 \pm 1.91$ & $14.97 \pm 1.87$ & $14.85 \pm 1.84$ \\ \hline
Levenshtein-Char & $16.63 \pm 1.98$ & $15.96 \pm 1.95$ & $15.56 \pm 1.90$ & $15.44 \pm 1.88$ \\ \hline
Levenshtein-Word & $14.98 \pm 1.87$ & $14.35 \pm 1.84$ & $13.97 \pm 1.79$ & $13.87 \pm 1.76$ \\ \hline
ROUGE-N          & $15.27 \pm 1.89$ & $14.59 \pm 1.86$ & $14.23 \pm 1.81$ & $14.08 \pm 1.79$ \\ \hline
ROUGE-L          & $15.46 \pm 1.90$ & $14.45 \pm 1.85$ & $14.09 \pm 1.80$ & $13.94 \pm 1.77$ \\ \hline
METEOR           & $14.45 \pm 1.82$ & $13.77 \pm 1.79$ & $13.42 \pm 1.74$ & $13.29 \pm 1.71$ \\ \hline
RoBERTa          & $14.80 \pm 1.86$ & $14.10 \pm 1.81$ & $13.76 \pm 1.77$ & $13.62 \pm 1.74$ \\ \hline
Cosine           & $14.99 \pm 1.88$ & $14.32 \pm 1.83$ & $13.99 \pm 1.79$ & $13.85 \pm 1.76$ \\ \hline
Euclidean        & $14.90 \pm 1.87$ & $14.24 \pm 1.82$ & $13.91 \pm 1.78$ & $13.78 \pm 1.75$ \\ \hline\hline
\textbf{Average (Mean)} & $\mathbf{15.23 \pm 1.89}$ & $\mathbf{14.53 \pm 1.85}$ & $\mathbf{14.17 \pm 1.80}$ & $\mathbf{14.06 \pm 1.77}$ \\ \hline
\textbf{StDev ($s$)}    & $0.6188$ & $0.6159$ & $0.5998$ & $0.5987$ \\ \hline
\textbf{Min}            & $14.45$ & $13.77$ & $13.42$ & $13.29$ \\ \hline
\textbf{Max}            & $16.63$ & $15.96$ & $15.56$ & $15.44$ \\ \hline
\end{tabular}
\end{scriptsize}
}
\end{table}

From the data presented in Table \ref{tab:AverageLLMQueries}, we observed that, first, in comparison with adversarial robustness evaluation methods, the proposed method requires far fewer LLM queries (fewer than 17 queries). Thus, it is much more efficient. Second, there is only a very small variation in the average numbers of queries to the LLMs required by different metrics. So, we have the following conclusion. 

\begin{observation}
From the point of view of the cost of using the LLM, the proposed method is feasible for practical use because it only requires a very modest number of queries to the LLM under test.  
\end{observation}

\subsubsection{CPU Time and Memory Space Consumptions} 

The required computation time for each test case consists of the time taken for each of the following steps. 
\begin{enumerate}
\item Generating a set of paraphrases. 
\item Calculating the distances between paraphrases and the original test case. 
\item Sorting the paraphrases according to distance.
\item Invoking the LLM to generate code from a paraphrase.  
\item Compiling and testing the generated code to determine if the LLM generated code is correct.
\end{enumerate}

The final two steps are repeated until incorrect code has been generated from a paraphrase by the LLM. 

From the analysis above, a key factor is the number of paraphrases generated from coding task description. In the experiments, we recorded the number of paraphrases generated from each test case, where the parameters are $n=5$ and $k=5$, which are about twice the average values of $n$ and $k$ required for tipping points; see Table \ref{tab:averageNandK}. 

\begin{figure}[htb]
\centering
\includegraphics[width=8cm]{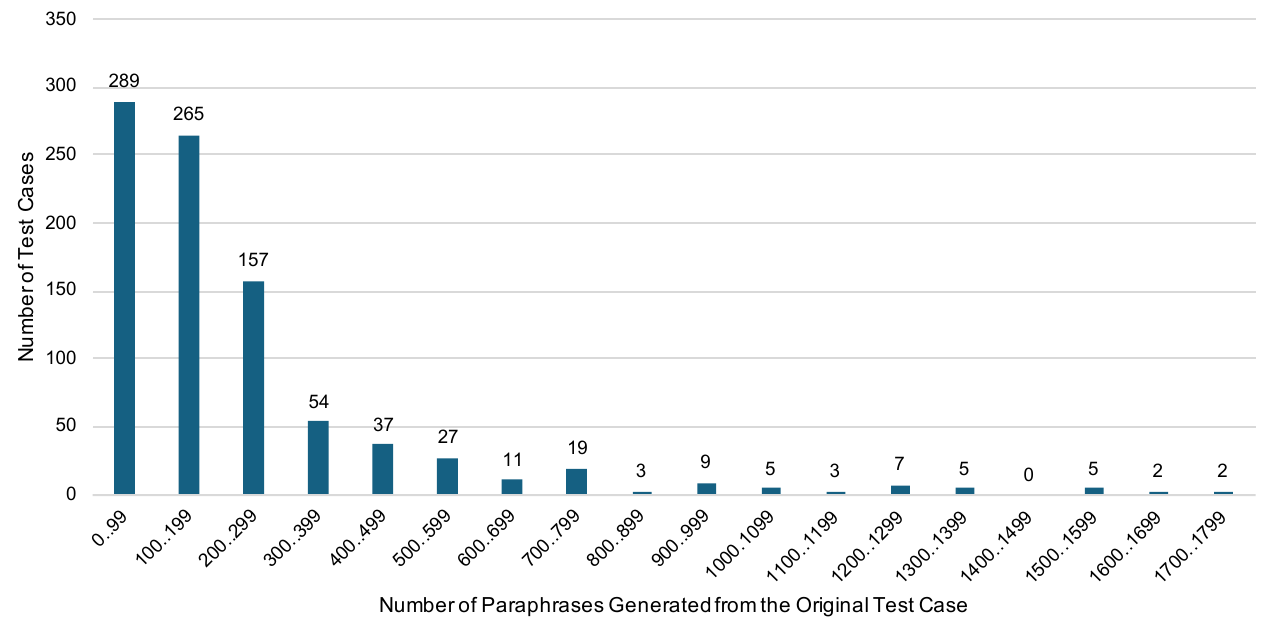} \\ 
\caption{Distribution of the Number of Generated Paraphrases}
\label{fig:Paraphrases}
\end{figure}

The experiment results are: the average number of paraphrases over all 900 test cases in the test dataset is 235.79, while the minimum and maximum number of paraphrases are 28 and 1770, respectively. The standard deviation is 269.74. Figure \ref{fig:Paraphrases} shows the distribution of the number of paraphrases generated from test cases. 

Therefore, we have the following observation. 

\begin{observation}
The generation of paraphrases and the sorting of them according to distance from the original coding task description can be very efficient. 
\end{observation}

We have also recorded the CPU time spent on calculating the distances (or similarity scores) between the paraphrases and the original coding task descriptions for each distance metric. Note that, given a distance metric, the time to calculate the distance between two texts may depend on the lengths of the texts. Table \ref{tab:distanceMetricTime} shows the average time  (in seconds) required to calculate one distance as well as the margin of error (MoE), the standard deviations, the minimum and maximum times.  

\begin{table}[ht]
\centering
\caption{Times to Calculate the Distance between The Original Task Description And A Paraphrase}
\label{tab:distanceMetricTime}
\resizebox{\linewidth}{!}{
\begin{tabular}{|l|c|c|c|c|c|}
\hline
\textbf{Metric} & \textbf{Average (ms)} & \textbf{MoE ($\pm$)} & \textbf{StDev ($s$)} & \textbf{Max} & \textbf{Min} \\ \hline
BLEU             & 0.2399 & 0.0153 & 0.0078 & 0.2922 & 0.2304 \\ \hline
ChrF             & 0.2489 & 0.0152 & 0.0078 & 0.2970 & 0.2394 \\ \hline
Levenshtein-Char & 0.4730 & 0.0164 & 0.0079 & 0.5229 & 0.4634 \\ \hline
Levenshtein-Word & 0.4799 & 0.0145 & 0.0079 & 0.5248 & 0.4704 \\ \hline
RougeN           & 0.8812 & 0.0161 & 0.0082 & 0.9333 & 0.8714 \\ \hline
RougeL           & 0.9119 & 0.0153 & 0.0078 & 0.9629 & 0.9024 \\ \hline
Meteor           & 1.1671 & 0.0161 & 0.0082 & 1.2213 & 1.1575 \\ \hline
RoBERTa          & 3.6302 & 0.0261 & 0.0082 & 3.6795 & 3.6205 \\ \hline
Cosine           & 3.7341 & 0.0251 & 0.0077 & 3.7751 & 3.7244 \\ \hline
Euclidean        & 3.9699 & 0.0354 & 0.0077 & 4.0172 & 3.9604 \\ \hline
\end{tabular}
}
\end{table}

From Table \ref{tab:distanceMetricTime}, we can observe that, first, the time required to calculate distances varies significantly among different metrics, ranging from 230 ms to nearly 4 seconds per paraphrase. Second, for a given distance metric, the variations on the times required to calculate the distances over the set of test cases is very small. Therefore, depending on the distance metric used, the time to calculate the distances takes from about 1 minute for BLEU and ChrF to about 16 minutes for RoBERTa, Cosine and Euclidean on average per test case. If the distance metric is one of RoBERTa, Cosine and Euclidean, calculating the distances for the set of paraphrases is one of the most time consuming steps. 

The required time for the final step consists of the time taken to query the LLM, generate test data from the generated code, compile and execute the generated code on the test cases, and determine the correctness of the generated code. Our experiment data shows that on average each query of the LLM may take about 18 seconds, while generation of test data takes about 320 seconds and the execution of the code takes about 16 seconds. In total, the final two steps take about 350 seconds. The generation of test data from the LLM generated code required the most computation time because it is implemented by employing the EvoSuite software testing tool, which is built on an evolutionary computing algorithm. 

The memory space consumed in the testing process consists mostly of the space used to store the set of paraphrases together with their distances to the original text. On average, less than 7MBs of memory space is required for each test case. However, when the test on one coding task completes, the memory space can be released. Therefore, the demand on memory space is minimal. 

From the above experiment data, we can now make the following conclusion. 

\begin{observation}
The computation time and memory space required to perform robustness test in our proposed method are affordable even when the most time consuming distance metrics are used. 
\end{observation}

It is worth noting that it is still time consuming to conduct a large scale experiment with thousands of test cases. However, experiments can be conducted by executing the algorithm on multiple desktop computers in parallel. 

\subsection{Threats to Validity}

This section discusses the threats to the validity of our experiments.  

\subsubsection{External Threats} 

Threats to external validity concern the extent to which our findings can be generalized. Our evaluation relies on the ScenEval benchmark, which exclusively consists of Java programming problems, including both text book and real-world Java programming tasks. However, the specific robustness metrics $R^o$ and $R^*$ for each LLM and in various scenarios may not be directly transferable from Java to other programming languages, as different languages have unique features. 
To address this limitation, future work could modify ScenEval to incorporate a broader range of languages or even use a different benchmark. However, despite this, we are confident that the conclusions we have drawn about the effectiveness and efficiency of the proposed method should be generalisable to all programming languages and all scenarios. In addition, the observations on the distinguishability of similarity/distance metrics should be generalisable since they are independent of the LLMs. 

In conclusion, the method proposed in this paper should be generalisable to other LLMs, programming languages and benchmarks. Nevertheless, further experiments to confirm this will be interesting.

\subsubsection{Internal Threats} 

One potential internal threat to the validity of our study is in the selection of test cases. If they are not sufficiently diverse, this could introduce bias in the results. For instance, over-representation of certain types of coding tasks may lead to misleading conclusions about the robustness of the LLMs. To mitigate this threat, we ensured that the test cases covered a broad range of topics and difficulty levels, ensuring a comprehensive evaluation. Moreover, we used random sampling where possible to avoid selection bias and to provide a more accurate assessment of the model performance across diverse scenarios. 

Another potential internal threat to the validity of our study is in the choices of the parameters of algorithm Alg. \ref{alg:ScenarioDomainAnalysis} in our experiments, which include $k$ and $n$ for generating the initial set of paraphrases, and $c_k$ and $c_n$ for generating additional paraphrases when an expansion of the set of paraphrases is required. These parameters have impact on the accuracy of the calculations of $R^o$ and $R^*$ metrics. The values of $k$ and $n$ used in our experiments were 5, which is about twice the average values of the tipping points (i.e. the last success and first failure points) over the whole test dataset. It is very rare that expansion is needed. The values of $c_k$ and $c_n$ were 1, which was sufficient to find the tipping points after expansion. 

To demonstrate the accuracy of our experimental evaluations of the robustness of LLMs, we refer to Theorem \ref{thm:Expansion2} below, which formally proves that the gaps between the values of $R^o$ and $R^*$ decreases monotonically with the increase of the mutant sample set. 

\begin{theorem}\label{thm:Expansion2}
Assume that both $Mut_t$ and $Mut'_t$ are valid sets of the mutants of $t$, $\left<l, f \right>$ and $\left<l', f' \right>$ are the pairs of last success and first failure points found in $Mut_t$ and $Mut'_t$, respectively. Then, $Mut_ \subseteq Mut'_t$ implies that 
\[l \preccurlyeq_t l'\preccurlyeq_t f' \preccurlyeq_t f,\] 
if $\|l, t\| \leq \nabla_{D^s}(t,M)$.\footnote{This condition normally holds when the set $Mut_t$ is generated with mutants closer to $t$ first as in Alg. \ref{alg:ScenarioDomainAnalysis}.}  \qed
\end{theorem}

Therefore, we can use the difference between the distances from the last success and first failure points to the seed test case as an indicator of the accuracy of the robustness evaluation. 
Table \ref{tab:Accuracy} gives the absolute differences between the experiment values of $R^o$ and $R^*$, which can be found in Table \ref{tab:combined_robustness}. Column \emph{Ratio} of Table \ref{tab:Accuracy} gives the ratios of the absolute differences between $R^o$ and $R^*$ to the mean values of them, i.e. 
\[Ratio = 2\frac{|R^o-R^*|}{R^o+R^*}. \]

The data in Table \ref{tab:Accuracy} shows that the overall average of the ratios is less than 5\%. This means the error rate in the experiment data is less than 5\%. Thus, the experiment data obtained from the experiment is highly accurate.  

\begin{table*}[htp]
\caption{Absolute Differences between $R^o$ and $R^*$ and The Ratios to Their Means}\label{tab:Accuracy}
\scriptsize
\begin{center}
\begin{tabular}{|c|c|c|c|c|c|c|c|c||c|c|}
\hline
\multirow{2}{*}{Similarity Metric} &\multicolumn{2}{|c|}{GeminiPro} &\multicolumn{2}{|c|}{Codex} &\multicolumn{2}{|c}{Llama2} &\multicolumn{2}{|c||}{Falcon} &\multicolumn{2}{|c|}{\textbf{Average}}\\\cline{2-11}
 &Difference &Ratio &Difference &Ratio &Difference &Ratio &Difference &Ratio &Difference &Ratio\\\hline
RoBERTa  &0.0752 &0.0700 &0.0553 &0.0496 &0.0533 &0.0436 &0.0504 &0.0410 &0.0586 &0.0511\\\hline
Euclidean  &0.0316 &0.0614 &0.0289 &0.0583 &0.0273 &0.0703 &0.0284 &0.0724 &0.0291 &0.0656\\\hline
Cosine Similarity &0.0011 &0.0012 &0.0080 &0.0086 &0.0010 &0.0010 &0.0009 &0.0009 &0.0027 &0.0029\\\hline
BLEU  &0.0292 &0.0661 &0.0224 &0.0524 &0.0292 &0.0533 &0.0288 &0.0530 &0.0274 &0.0562\\\hline
ROUGE-N  &0.0204 &0.0322 &0.0233 &0.0360 &0.0248 &0.0329 &0.0225 &0.0300 &0.0228 &0.0328\\\hline
ROUGE-L  &0.0180 &0.0256 &0.0169 &0.0236 &0.0173 &0.0211 &0.0157 &0.0192 &0.0170 &0.0224\\\hline
METEOR  &0.0233 &0.0349 &0.0233 &0.0342 &0.0242 &0.0307 &0.0221 &0.0282 &0.0232 &0.0320\\\hline
ChrF  &1.8729 &0.0276 &1.7555 &0.0255 &1.6484 &0.0237 &1.7166 &0.0248 &1.7484 &0.0254\\\hline
Levenshtein-Word  &0.1110 &0.1029 &0.1137 &0.1061 &0.1164 &0.1090 &0.0697 &0.0639 &0.1027 &0.0955\\\hline
Levenshtein-Char  &1.4018 &0.1296 &1.0976 &0.1093 &0.9640 &0.0999 &1.0320 &0.1051 &1.1239 &0.1110\\\hline\hline
\textbf{Average} &\textbf{0.3585} &\textbf{0.0551} &\textbf{0.3145} &\textbf{0.0504} &\textbf{0.2906} &\textbf{0.0486} &\textbf{0.2987} &\textbf{0.0439} &\textbf{0.3156} &\textbf{\textit{0.0495}}\\\hline
\end{tabular}
\end{center}
\label{default}
\end{table*}%

\subsubsection{Construct Threats}

Construct validity concerns the appropriateness of the measures and metrics used in the experiments. It manifests itself in the way we determine the functional equivalence between the codes generated from the original coding task and from its paraphrases, and in the metrics used to order the mutants for testing. 

Given the fact that the problem whether two programs are functionally equivalent is not computational decidable, we take the practical approach of test equivalence, which is the approach widely adopted by researchers on code generation. However, we have strengthened the approach by requiring two conditions: (a) the generated code must pass all test cases generated from the reference program, and (b) the reference program must pass test cases generated from the generated code. The former enables us to detect omission errors, while the latter enables us to detect commission errors. In the current research on code generation, only condition (a) is applied, thus it is unable to detect commission errors. Therefore, by employing both (a) and (b) in this paper, we have reduced the potential threat to construct validity by reducing the risk that the test suite may be inadequate and that the errors in the generated programs might be overlooked. To further mitigate this threat, we implemented the automatic test case generation by utilising the EvoSuite tool \citep{fraser2014, fraser2011} as detailed in \citep{ghosh2024}. EvoSuite is perhaps the best white-box test case generation tool. It is well known for its strong test adequacy \citep{fraser2014, fraser2011}, ensuring generating high test coverage of the code. This is also proved empirically in our previous work \citet{ghosh2024}. Consequently, there is minimal risk that the results of the experiment can have systematic bias due to omission of critical program bugs. 

Another threat to validity is that the wrong distance metric was used to order the mutants. Therefore, we studied a wide range of metrics covering all types of such metrics. 
For all of these metrics, the experiments demonstrated that the proposed method can provide meaningful evaluations with high precision consistently while different similarity metrics have different capability of distinguishing LLMs. The experiments also demonstrated the feasibility of the proposed method. While these metrics revealed subtle differences in the results of the evaluation of LLM's robustness on code generation, the impact of the choice of distance metrics is minimal due to the consistency in their results. Thus, the risk associated with the choice is minimal. 

\section{Related Work}\label{sec:RelatedWork}

In this section, we review the related works and compare them with the method proposed in this paper. We will limit our scope to testing and evaluation of robustness. The works on defence against adversarial attacks and improvement of robustness are omitted. 

We will first discuss the relationship between the notion of micro operational robustness proposed in this paper and the notion of adversarial robustness in the research on ML models. Then, we review works on robustness of code generation. We extend the scope of review to robustness of NLP because its input space is the same as that of code generation. 

\subsection{Relationships with Adversarial Robustness}

Our notion of micro operational robustness stems from and generalises the notion of \emph{adversarial robustness} based on Szegedy et al.'s notion of adversarial examples \cite{szegedy2014}. Given a ML model $M$ and $x$ in the input space $D$ of $M$, $x' \in D$ is an \emph{adversarial example} (AE) of $x$ if $M(x) \neq M(x')$ and the difference between $x$ and $x'$ is not perceptible by human beings, which is formalised and generalised as $\|x,x'\| < \epsilon$ for a given very small number $\epsilon >0$. 

In the past two decades, the research on testing and evaluation of adversarial robustness have been dominated by developing algorithms that calculate and/or search for adversarial examples of ML models. Notably, such attack models can be classified into white-box and black-box attacks depending on whether the details of the ML model is used to generate AEs \cite{khamaiseh2022,tekeste2026adversarial}. 
The \emph{average distortion} introduced by Szegedy et al. \cite{szegedy2014} are widely used by many researchers to measure ML model $M$'s adversarial robustness, e.g., \citet{fawzi2016robustness, fawzi2018analysis}. 
\begin{equation}
R_{G}(T,M) = \frac{1}{|T|} \sum_{x \in T} \|x,G(x,M) \| \label{equ:robustnessInAttackModel}
\end{equation}
where $T$ is the benchmark of test cases from which adversarial examples were generated. $G(x,M)$ is an algorithm that generates adversarial examples in an attack model $\alpha$. 

To be a theoretically sound measurement of robustness, Moosavi-Dezfooli \emph{et al.} \citet{moosavi2016} further requires the AEs generated by an algorithm $G(x,M)$ to be minimal. That is 
\[\|x,G(x,M)\|=\min_{x' \in D^\alpha}\{\|x,x'\| ~|~ M(x) \neq M(x')\}.\] 
where $D^\alpha$ is the set of all possible adversarial attacks in the attach model $\alpha$.
Let $\Delta_{D^\alpha}(x,M)$ denote the minimal AE of $x$ for ML $M$. Formally, 
\[\Delta_{D^\alpha}(x,M) = \min_{x' \in D^\alpha}\{\|x,x'\| ~|~ M(x) \neq M(x')\}.\] 
$R_{G}(T,M)$ in Equ. (\ref{equ:robustnessInAttackModel}) is actually a statistical estimation of the value of $\rho_\alpha$ below.
\begin{equation}
\rho_\alpha(M)= E_{x}\left(\Delta_{D^\alpha}(x,M)\right) \label{equ:AdRobustnessAttackModel}
\end{equation}

Our definition of micro operational robustness can be regarded as replacing the set $D^\alpha$ of AEs in attack model $\alpha$ in Equ (\ref{equ:AdRobustnessAttackModel}) with the set $D^s$ of input data in scenario $s$. Therefore, 
our definition is consistent with the notion of adversarial robustness. 

Informally, in Equ (\ref{equ:AdRobustnessAttackModel}), $\Delta(x,M)$ gives the size of the safe zone that surrounds the input $x$ in which adversarial attacks will not succeed; while outside the zone, the ML model may fail under attack. Thus, $\Delta(x,M)$ gives the boundary between the safe and unsafe zones. In this paper, we discovered that there may be a gap between the safe zone and unsafe zone. The exact boundaries of them are not necessarily equal in discrete and sparse input spaces; while they are equal for continuous and dense input spaces like images. Thus, our definition of robustness with two values $\nabla(x,M)$ and $\Delta(x,M)$ for the sizes of safe and unsafe zones, respectively, generalises the definition so that it is applicable to all type of input spaces. 

\subsection{Robustness of Code Generation}

The only work belonging to the white-box approach to testing and evaluation of the adversarial robustness of code generation is perhaps 
Zhu and Zhang's work in 2023 \cite{zhu2023robust}. They took a white-box approach to test and evaluate GPT-2 on its robustness when generating Python code. They defined a loss function based on the logit layer of the LLM and the output labels of each token in the output code. The loss function is used to guide the perturbations on the input coding task description and then iteratively updated for the generation of next input through back-propagation until an adversarial example is generated. The oracle that decides the termination of the iterations is the BLEU score between the output on the original coding task and the output on the perturbed input. Their experiment used a subset of the APPS benchmark with difficulty of coding tasks at three levels: introductory, interview and competition levels. They measured the effectiveness of attacks by two metrics: (a) TED: the editing distance between the abstract syntax trees; (b) ASR: the attack success rate, which was defined as the proportion of the code generated by the adversarial examples that succeeds over all adversarial examples generated. Success for an adversarial example is defined as the output code generated from the adversarial input coding task passes at least one test case. Their results are that the average TED and ASR over all levels of coding task difficulty were 34.20\% and 64.99\%, respectively. They also measured the quality of the adversarial examples by the average BLEU score and number of words changed, where were 97.90 and 1.8, respectively. 

Zhu and Zhang's work \cite{zhu2023robust} differs from our approach in that they have taken a white-box approach while our approach is black-box. White-box approach is not applicable to LLMs that are not open. Moreover, it is unclear how efficient is Zhu and Zhang's approach. It could be very expensive to run as the LLMs are huge in scale.  

In fact, most of the works on robustness of code generation belong to the black-box approach. 
The earliest one on the adversarial robustness of code generation is perhaps by Anand and Kayal in 2021 \cite{anand2021adversarial}. They studied the SketchAdapt (SA) model and compared it with a transformer model (TF) on code generation from textual descriptions. Five types of adversarial attacks were implemented as transformations of the textual descriptions. These are: 
(a) variable change, 
(b) redundancy removal, 
(c) synonym replacement, 
(d) voice conversion, and 
(e) variable interchange. 
These perturbations are applied to the textual description of coding tasks in the AlgoLisp dataset and generated about 1000 adversarial instances. The error rates of the SA and TF models on each type of attacks were calculated. They calculated the \emph{extents of perturbations} of the adversarial instance for each type of attacks using edit distance and embedding similarity, and the quality of the adversarial instances on grammatical correctness and naturalness via  manual evaluation. However, in contrast to our approach, there is no effort to find the minimal adversarial instances. 

Liguori et al. \cite{liguori2022can} are perhaps the earliest researchers addressed the operational robustness of code generation. They were concerned with the sensitivity to variations of NL descriptions of coding tasks from well-intentioned users rather than adversaries. Three types of perturbations were used to generate modified descriptions of coding tasks from a dataset of 305 test cases that each consists of assembly instructions commented in English. They evaluated one ML model on the original test dataset and three perturbed test datasets constructed using the following three perturbations. 

(a) Unseen synonyms with constraints using BERT-score and part-of-speech (POS) tag: a word in the description is replaced by a synonym chosen from the neighbours in the word embedding space under the constraint that the synonym and the original word have a BERT similarity score $\geq 0.85$ and have the same POS tag. With in a NL description, 10\% of the words were selected at random to be replaced by such a synonym and at least one word is replaced. 

(b) Omission of action-related words: the verbs in the NL description is replaced by the POS tagger, e.g. "define", "add", etc. 

(c) Omission of language related words:  the words related to the assembly programming language in the NL description are removed. 

The correctness of the output on a perturbed input is determined by comparing with the reference solution using metrics BLUE-4, exact match accuracy, editing distance, and the LCS-based metric in instead of functional testing the generated code. The syntax and semantic correctness of generated codes were also manually evaluated. 

Liguori et al.'s perturbations \cite{liguori2022can} were employed in their recent work on testing LLM vulnerability to NL perturbations and improving their robustness \cite{improta2025enhancing}. They used these perturbations as data augmentations to generate synthetic training data to fine tune LLMs. 

Shirafuji et al. explored the robustness of LLMs on solving programming problems described in NL but in a specific format that contains a specification of the functional requirements, descriptions of input and output together with sample inputs and outputs \cite{shirafuji2022prompt,shirafuji2023exploring}. In addition to randomized replacement words with synonyms, they applied three more modifications on such problem descriptions that were specific to programming problem specifications. These are: (a) randomized change of variable names occurring in the problem description, (b) anonymizing output strings, (c) modifying the algorithmic and computing terminology in the problem specification to the opposite but still valid, such as changing "ascending order" to "descending order". Each modification operator is applied to the problem specification 20 times and the performance of the Codex is evaluated on the original and modified problem specifications. 

Inspired by works on robustness of NLP, Zhuo et al. \cite{zhuo2023robustness} used 5 word level perturbations and 2 sentence level perturbations in a case study with Codex. They applied these perturbations on a benchmark dataset to form perturbed datasets and evaluated the performances of Codex called \emph{standard accuracy} on the original dataset, the \emph{perturbation accuracy} on the perturbed dataset, and also \emph{robust accuracy}, which is $n/r$ , where $r$ is the number of test cases in $R_{eval}$, the subset of perturbed dataset on which the LLM is correct on their original input, $n$ is the number of test cases in $R_{eval}$ that the LLM is also correct. In other words, robust accuracy equals 1 minus attack success rate. 

Mastropaolo et al. conducted an empirical study of the robustness of code generation with GitHub Co-Pilot \cite{mastropaolo2023}. They implemented two transformations on NL descriptions of coding tasks: (a) a deep learning based paraphrasing tool PEGASUS \citep{zhang2020pegasus}, and (b) translation pivoting (TP) technique, i.e. translation from English to French and then back. 

Yan et al. \citet{yan2024robustness} proposed a technique called COCO in which the coding task is augmented with features extracted from the code generated for the original task description. 

More recently, in a study of the impact of quantization on LLM robustness of code generation, Fang et al. employed Liguori et al.'s word substitution perturbation  \cite{fang2025smaller}. In addition to this, they also used a character level perturbation, which randomly selects a character and converts it into upper case, and a sentence level one, which translates a sentence into Germany and then back to English. For each of these perturbations, a set of attacks were constructed and used to test the LLMs and their quantized versions. The performances of the models are measured by correctness in pass@1. 

Despite the differences in the perturbations developed and used by researchers in the study of the robustness of code generation, a common characteristic of the above works is that robustness was measured by macro metrics like attack success rate. Therefore, they all belong to the macro robustness approach. In contrast, our method uses the micro approach. 

Moreover, the word substitution with synonyms has been used by many researchers although there are subtle differences in how synonyms were selected and applied. It is very similar to our uses of synonym to perturb coding task descriptions. However, the key difference is that they made no effort to find the minimal variations of all possible ones. 

\subsection{Adversarial Robustness of NLP}

There is a vast number of works on adversarial robustness of ML models for image processing. However, as Yefet \emph{et al.} pointed out \citet{yefet2020adversarial}, such algorithms that work well on continuous and dense input spaces like image spaces cannot easily be adapted to NLP models, which are discrete and sparse. Therefore, we will omit them in our discussion of related works. Readers are referred to \cite{khamaiseh2022,tekeste2026adversarial} for recent surveys of research on adversarial robustness in image processing. 

In the research on adversarial robustness of ML on NLP, the white box approaches include Ebrahimi \emph{et al.}'s \emph{HotFlip}, which \citet{ebrahimi2018} adapted Goodfellow's FGSM algorithm for image processing \citet{goodfellow2014} to identify the most influential characters in a word based on the model's gradients. Adversarial examples are generated by flipping those characters. \emph{TextFool} proposed by Liang \emph{et al.} \citet{liang2018} changes entire words rather than characters and thereby achieves more impactful modifications.  \emph{TextBugger} proposed by Li \emph{et al.} \citet{li2019textbugger} is a JSMA-based algorithm \citep{papernot2016} that manipulates specific words in the text by calculating the Jacobian matrix. It then applies subtle perturbations, like inserting spaces, deleting letters, replacing letters, or swapping adjacent letters to reduce the model's confidence in prediction.

\emph{Black-box approaches} generate AEs without requiring detailed knowledge of the ML model's internal structure or parameters. Such information is usually not available to the attackers. Instead, the attacker can only see the output labels and associated confidence scores. Such information is used to search for the minimal (in most cases, the local minimal) perturbations that causes the ML model to fail. 

One of the earliest works on the black-box approach to adversarial robustness for NLP is the work by Jia \emph{et al.} \citet{jia2017} and Wang et al. \citet{wang2018robust} on the robustness of ML models for reading comprehension. They explored a variety of methods of perturbations of the text to be understood by adding a sequence of words, which include constructing a sentence that is syntactically identical but semantically different from original sentence to be understood, adding a random sequence of words drawn from the Brown corpus of 1000 most common English words and the list of words in the original sentence to be understood, etc.

The score-based black-box approach to the generation of adversarial examples modifies various words or phrases in the input text and compare the scores (e.g. confidence) of ML model's prediction before and after the change. For example, in their research on the robustness of sentiment analysis and topic classification, Gao \emph{et al.} \cite{gao2018blackbox} proposed the \emph{DeepWordBug} algorithm which ranks words based on their importance to the model's prediction and applies character-level perturbations (e.g., insertion, deletion, substitution, or swapping). 
\emph{PWWS} (Probability Weighted Word Saliency) introduced by Ren \emph{et al.}\citet{ren2019pwws} uses perturbations at the word level. They score words based on their contribution to the 
prediction and replace high-saliency words with their synonyms iteratively to reduce the model's confidence. 

Also within the black-box approach, there is much work on fuzzing testing of NLP models. They are not only applicable for evaluating robustness, but a more general automatic test case generation technique. The basic idea is to make random modifications on the input by applying a set of predefined perturbations, which are also called transformations in the literature. Typically, fuzzing is performed iteratively according to a certain strategy. The goal is to search for the input that cause the ML model to fail. It can be applied to robustness testing and evaluation when failure is defined as generating a different output from the original output. Two types of strategies have been advanced. 

\emph{Static strategies} generate a set of perturbed inputs without dynamically invoking the LLM under test. Typical examples of this approach include the works by Wang et al. on metamorphic testing of ML models on content moderation \cite{wang2023mttm}, Yu et al.'s work on testing of DNNs on names entity recognition \cite{yu2023automated}, Wang et al.'s work on testing the robustness of ChatGPT \cite{wang2023robustness, wang2024robustness}, and Yuan et al.'s work on out-of-distribution robustness of LLMs on 5 different NLP tasks \cite{yuan2023revisiting}. This category of works measures robustness by the ML model's failure rate. Thus, they belong to macro robustness approach and hence differs from our approach. 

\emph{Dynamic strategies} iteratively applies perturbations to the input and test the ML model to decide if further applications will be applied according to, for example, whether the ML model fails and if the complexity of the input is too high.  Complex strategies have been developed to maximise the success rate of exposing failures of the software under test and minimise the time required to find failure exposing inputs. For example, Wang et al. utilised a genetic algorithm \cite{wang2021natural}. The LEAP method proposed by Xiao et al. employs swarm optimisation algorithm to improve search efficiency \cite{xiao2023leap}. A recent work that represents the current state of art in this approach is Xiao et al's BASFuzz method, which employs the beam search algorithm with an annealing mechanism \cite{xiao2025basfuzz}. In this approach, ML models are compared by their failure rates (or the attack success rate) while the quality of the attacks are measured by the similarity or distance of the adversarial examples to the original inputs. Again, there is no attempt to search for the minimal adversarial examples according to a distance metric. This is the key difference from our method. Although great efforts have been made by many researchers to improve the efficiency of such testing methods, the cost of such algorithms is very high. For example, the current state of art as represented by BASFuzz requires 148 to 286 queries of LLMs and 284 to 553 seconds to generate one adversarial example depending on the dataset \cite{xiao2025basfuzz}.

\section{Conclusion}\label{sec:Conclusion}

\subsection{Main Contributions} 

The main contributions of the paper are as follows. 

First, we proposed the notion of micro operational robustness of ML models in operational scenarios. The notion is formally defined as the expected minimal distance to failure and maximal safe distance in the operation scenario from a data in the input space over the data distribution. The properties of micro operational robustness were studied. We discovered that, for discrete and sparse input spaces like NL and program code, there may be a gap between the minimal distance to failure and maximal safe distance while they are equal in continuous and dense input spaces like images. 

Secondly, we proposed an algorithm to test and evaluate the micro operational robustness in operational scenarios for discrete and sparse input spaces. We formally proved the algorithm's correctness with respect to the formally defined notion of micro operational robustness under the assumption that the mutant generation function invoked in the algorithm satisfies validity and relative completeness conditions. These conditions can be validated informally against the understanding of the operational scenarios. 

Moreover, we applied the proposed algorithm to LLM on code generation. It is the first method for evaluating micro operational robustness of NLP on code generation and also on other generative tasks. Through experimentation, we have demonstrated that the proposed method is both feasible, effective and efficient for evaluating the robustness of LLM models. It requires fewer than 20 queries to the LLM per test case and the required computation resources in terms of CPU time and memory are very modest and affordable. 

Finally, we conducted intensive experiments with LLMs on code generation with a wide range of distance/similarity metrics on inputs, i.e. coding task descriptions in NL. Our experimental results offer valuable insights into the robustness of LLMs on code generation. In particular, in terms of the similarity between the original coding task description and the paraphrases, we found that an LLM, like Codex, will start to generate incorrect programs from the paraphrase when the paraphrase similarity is 0.42 for BLEU,  0.93 for Cosine, and 0.66 for Meteor, for example. This means, roughly speaking, on average a state-of-the-art LLM will start to generate incorrect programs if the task description is modified by replacing 2 to 3 words in the text with words of similar meanings (the 3 to 4 closest ones in word embedding vector space). Moreover, we have studied the robustness in various sub-scenarios in various coding topics and coding task complexities. Our experiment data shows that, for all LLMs, robustness is weaker in advanced coding topics such as multi-threading and data structures than in basic topics like array and string processing. And, robustness is highly correlated to coding task complexity. It decreases significantly with the increase of coding task complexity. 

\subsection{Future work}

We have implemented a test system that automates the robustness evaluation process in our proposed method with the support of the Morphy tool \cite{zhu2020}. Due to the space limitation, it is not included in this paper. It will be reported separately. 

As one would expect, the definition of micro operational robustness depends on the distance/similarity metric on the input space. Our experiments also demonstrated that the robustness evaluation result are influenced by such metrics. A further research question is how to select the metric. We have conducted a study of the distance/similarity metrics in the context of micro robustness \cite{PaulPhDThesis}. The work will also be reported in a separate paper. 

LLMs evolve quickly with new versions and new models being announced frequently. It will be interesting to find how their robustness in code generation evolves. 

In this paper we use test correctness as the test oracle to determine the robustness of code generation. Test oracles can also be defined, for example, based on distance/similarity of the generated code with respect to the original code. Aspects of the quality of the generated code, such as code smells, could also be used to define test oracles. By analysing these kinds of robustness, we can provide a more comprehensive understanding of LLM robustness in code generation, ultimately leading to insights that can enhance the usability and effectiveness of code generation in real-world applications.

Finally, the proposed theoretical framework of micro operational robustness and the algorithm for discrete and sparse input space can be applied to a wide range of ML application domains, including all NLP tasks. Moreover, LLMs have also been widely used for many other types of coding tasks such as clone detection, debugging or code repairing, code refactoring, code summary and comment generation, vulnerability detection, etc. Their robustness has been a hot topic, but all existing works have taken the macro approach \cite{qu2024survey}. The input space of such coding tasks is program code, which is also discrete and sparse. Therefore, the proposed method should be applicable and it will be very interesting to see this done.



\begin{footnotesize}

\end{footnotesize}

                                  
\newpage

\section*{Appendix A. Proofs of Theorems and Lemmas} 

This appendix gives the formal proofs of the Lemmas and Theorems presented in the main text of the paper. For the sake of self containment, we repeat here the definitions and the statements of the lemmas and theorems. 

\vspace{+10pt}
\noindent{\textbf{\textit{Definition} \ref{def:MinimalDistanceToFailure}.} (Minimal distance to failure in a scenario)}

Let $t \in D$ and $D^s \subset D$ be the subset of the input space representing an operational scenario $s$.
 
An ML model $M$'s \emph{minimal distance from $t$ to failure} in the scenario $s$, denoted by $\Delta_{D^s}(t,M)$, is defined as follows. 

\begin{equation}
\Delta_{D^s}(t,M)=\min_{x \in D^s}\{\|x,t\| ~|~ Fails^M_t(x)\}. 
\end{equation}
The data $x\in D^s$ such that $\|x,t\|=\Delta_{D^s}(t,M)$ is called a \emph{minimal deviation} of $t$ in scenario $s$. 
\qed

\vspace{+10pt}
\noindent{\textbf{\textit{Lemma} \ref{lmm:lmmA}.}}

For all $x \in D^s$, we have that 
\[\left(\|x,t\| < \Delta_{D^s}(t,M) \Rightarrow \neg Fails^M_t(x)\right).\] 

\proof{Lemma \ref{lmm:lmmA}} 

Directly from Definition \ref{def:MinimalDistanceToFailure}. 
\qed

\vspace{+10pt}
\noindent{\textbf{\textit{Definition} \ref{def:Safe}.} (Safe mutants)}

Let $t \in D$. We say that $x \in D^s$ is \emph{safe} for model $M$ w.r.t. to $t$, denoted by $Safe^M_t(x)$, if and only if (a) the model $M$ does not fail on $x$, and (b) $M$ does not fail on any points $x'$ in $D^s$ whose distance to $t$ is less than the distance from $x$ to $t$. Formally, 
\begin{eqnarray}
\lefteqn{Safe^M_t(x) = \neg Fails^M_t(x)~ \wedge }\nonumber \\
&& \forall x' \in D^s.( \| x', t\| < \| x,t\| \Rightarrow \neg Fails^M_t(x'))
\end{eqnarray}
\qed

\vspace{+10pt}
\noindent{\textbf{\textit{Definition} \ref{def:SafeDistance}}. (Maximal safe distance in a scenario)}

The \emph{maximal safe distance} from $t$ in a scenario domain $D^s$ for ML model $M$, denoted by $\nabla_{D^s}(t,M)$, is defined as follows. 
\begin{equation}
\nabla_{D^s}(t,M)=\max_{x \in D^s}\{\|x,t\| ~| Safe^M_t(x)\}. 
\end{equation}
\qed

\vspace{+10pt}
\noindent{\textbf{\textit{Lemma} \ref{lmm:lmmC}.}}

$\forall x \in D^s.\left(\|x,t\| < \nabla_{D^s}(t,M) \Rightarrow Safe^M_t(x)\right).$ 

\proof{Lemma \ref{lmm:lmmC}}

We prove this by contradiction. Assume that the statement of the lemma is not true, i.e. $x \in D^s$ and $\|x,t\| < \nabla_{D^s}(t,M)$, but $\neg Safe^M_t(x)$. 

Let 
\begin{equation}
\alpha = \underset{x \in D^s}{\operatorname{argmax}} \left(\{\|x,t\| ~| Safe^M_t(x)\}\right) \label{eqn:proof20}
\end{equation}
By Definition \ref{def:SafeDistance}, $\|x,t\| < \nabla_{D^s}(t,M)$ implies that  
\begin{equation}
\|x,t\| < \|\alpha, t\|. \label{eqn:proof21}
\end{equation}

By Definition \ref{def:Safe}, $\neg Safe^M_t(x)$ implies that $Fail^M_t(x)$ or $\exists x' \in D^s.\left( \| x', t\| < \| x,t\| \wedge Fails^M_t(x')\right)$. The following proves that both cases are not true. 

Case (1): $Fail^M_t(x)$. 

By Equ. (\ref{eqn:proof20}), we have that $Saft^M_t(\alpha)$. By Definition \ref{def:Safe}, Equ. (\ref{eqn:proof21}) implies that $\neg Fail^M_t(\alpha)$. Therefore, Case (1) is not true. 

Case (2): $\exists x' \in D^s.\left( \| x', t\| < \| x,t\| \wedge Fails^M_t(x')\right)$. 

Let $x' \in D^s$ such that $\| x', t\| < \| x,t\| \wedge Fails^M_t(x')$. Equ. (\ref{eqn:proof21}) implies that $\|x',t\| < \|\alpha, t\|$. Again, by Definition \ref{def:Safe}, this implies that $Fails^M_t(x')$ is not true. Therefore, Case (2) is not true. 

Because both cases (1) and (2) are not true, the assumption is not true. The statement of the lemma holds. 
\qed

\vspace{+10pt}
\noindent{\textbf{\textit{Lemma} \ref{lmm:lmmD}.}}

$\forall x \in D^s.\left(Safe^M_t(x) \Rightarrow \|x,t\| \leq \nabla_{D^s}(t,M)\right)$. 

\proof{Lemma \ref{lmm:lmmD}}

We prove the lemma by contradiction, again. Assume that $x \in D^s$ such that $Safe^M_t(x)$ and $\|x,t\| > \nabla_{D^s}(t,M)$. By Definition \ref{def:SafeDistance}, this implies that 
\[
\|x,t\| > \max_{x' \in D^s}\{\|x',t\| ~| Safe^M_t(x')\}
\]
This contradicts $Safe^M_t(x)$. Therefore, the lemma is true. 
\qed

\vspace{+10pt}
\noindent{\textbf{\textit{Theorem} \ref{thm:thm1}.}}

$\forall t \in D. \left(\nabla_{D^s}(t,M) \leq \Delta_{D^s}(t,M)\right)$. 

\proof{Theorem \ref{thm:thm1}} 

We prove the theorem by contradiction. 

Assume that there is a $t \in D$ such that $\nabla_{D^s}(t,M) > \Delta_{D^s}(t,M)$. Let 
\begin{eqnarray}
\alpha = \underset{x \in D^s}{\operatorname{arg}}\{\|x,t\|=\nabla_{D^s}(t,M) \wedge Safe^M_t(x)\} \label{eqn:proof22}\\
\beta = \underset{x \in D^s}{\operatorname{arg}}\{\|x,t\|=\Delta_{D^s}(t,M) \wedge Fail^M_t(x)\} \label{eqn:proof23}
\end{eqnarray}

By Definition \ref{def:MinimalDistanceToFailure} and \ref{def:SafeDistance}, $\alpha$ and $\beta$ exist because the conditions $Safe^M_t(x)$ and $Fail^M_t(x)$ in Equ. (\ref{eqn:proof22}) and (\ref{eqn:proof23}) are already a condition of $\nabla_{D^s}(t,M)$ and $\Delta_{D^s}(t,M)$, respectively. The definition of $\alpha$ and $\beta$ imply that $Safe^M_t(\alpha)$ and $Fail^M_t(\beta)$. 

Assumption $\nabla_{D^s}(t,M)>\Delta_{D^s}(t,M)$ implies that $\|\alpha,t\|$ $>\|\beta, t\|$. By Definition \ref{def:Safe}, this means $\neg Safe^M_t(\alpha)$, because $Fail^M_t(\beta)$. This contradicts the assertion that $Safe^M_t(\alpha)$. 

Therefore, the assumption is not true; hence the theorem holds. 
\qed

\vspace{+10pt}
\vspace{+10pt}
\noindent{\textbf{\textit{Theorem \ref{thm:NablaDelta}.}}}

There is no $x \in D^s$ such that $\nabla_{D^s}(t,M) < \|x, t\| < \Delta_{D^s}(t,M)$. 

\proof{Theorem \ref{thm:NablaDelta}}  

We prove the theorem by contradiction. Assume the theorem is not true; that is, there is $\alpha \in D^s$ such that 
\begin{equation}
\nabla_{D^s}(t,M) < \|\alpha, t\| < \Delta_{D^s}(t,M). \label{enq:thm0eqn1}
\end{equation}

We have that either $Fail^M_t(\alpha)$ or $\neg Fail^M_t(\alpha)$. The following proves that in both cases, the assumption (\ref{enq:thm0eqn1}) is not true. 

\emph{Case} (1): Assume that 
\begin{equation}
Fail^M_t(\alpha). \label{eqn:assumption1}
\end{equation}

By Lemma \ref{lmm:lmmA}, from (\ref{eqn:assumption1}) we have that $\|\alpha,t\| \geq \Delta_{D^s}(t,M)$. This contradicts (\ref{enq:thm0eqn1}). Therefore, assumption (\ref{eqn:assumption1}) is not true. 

\emph{Case} (2): Assume that 
\begin{equation}
\neg Fail^M_t(\alpha). \label{eqn:assumption2}
\end{equation}
By (\ref{enq:thm0eqn1}), i.e. $\|\alpha, t\| > \nabla_{D^s}(t,M)$, and Lemma \ref{lmm:lmmD}, we have that 
\begin{equation}
\neg Safe^M_t(\alpha). \label{eqn:thm0eqn1}
\end{equation}
By Definition \ref{def:Safe}, (\ref{eqn:assumption2}) and (\ref{eqn:thm0eqn1}) imply that there is $x \in D^s$ that $(\|x,t\| < \|\alpha, t\|) \wedge Fail^M_t(x)$. By (\ref{enq:thm0eqn1}), this implies that $\|x,t\| < \Delta_{D^s}(t,M) \wedge Fail^M_t(x)$. It contradicts Lemma \ref{lmm:lmmA}. Therefore, assumption (\ref{eqn:assumption2}) is not true. 

Because both (\ref{eqn:assumption1}) and (\ref{eqn:assumption2}) are not true, we have that the assumption about the existence of $\alpha$ that satisfy (\ref{enq:thm0eqn1}) is not true. Thus, the theorem holds. 
\qed

\vspace{+10pt}
\noindent{\textit{\textbf{Lemma \ref{lmm:lmm1}.}}}

$\forall x \in Mut_t.\left(Fail^M_t(x) \Rightarrow (FF(t) \preccurlyeq_t x)\right)$. 

\proof{Lemma \ref{lmm:lmm1}}

Let $x \in Mut_t$ and $Fail^M_t(x)$. If $x \prec_t FF(t)$, $x$ will be ordered before $FF(t)$. $Fail^M_t(x)$ contradicts that $FF(t)$ is the first failure. 
\qed

\vspace{+10pt}
\noindent{\textit{\textbf{Lemma \ref{lmm:llm2}.}}}

$LS(t) \preccurlyeq_t FF(t)$. 
    
\proof{Lemma \ref{lmm:llm2}}

By definition of last success and first failure, $LS(t)$ is always in the order before $FF(t)$. 
\qed

\vspace{+10pt}
\noindent{\textit{\textbf{Lemma \ref{lmm:lmm3}.}}}

$\neg \exists x \in Mut_t.\left(LS(t) \prec_t x \prec_t FF(t)\right)$. 

\proof{Lemma \ref{lmm:lmm3}}

Assume that $\alpha \in Mut_t$ and 
\begin{equation}
LS(t) \prec_t \alpha \prec_t FF(t). \label{eqn:lmm2Proof1}
\end{equation}

If $Fail^M_t(\alpha)$, by Lemma \ref{lmm:lmm1}, we have that $FF(t) \preccurlyeq \alpha$. This contradicts (\ref{eqn:lmm2Proof1}). 

If $\neg Fail^M_t(\alpha)$, (\ref{eqn:lmm2Proof1}) contradicts that $LS(t)$ is the last success. 

Therefore, $\alpha \in Mut_t$ is not true. Thus, the Lemma statement holds. 
\qed


Let $K>0$ and $mut_1, \cdots, mut_K$ be sets of mutants of test case $t$ such that 
\begin{enumerate}
\item $mut_i \cap mut_j = \emptyset$, if $i\neq j$; 
\item for all $x \in mut_i$, $\neg Fail^M_t(x)$, for all $i = 1, \cdots, K-1$; 
\item there exists an $x \in mut_K$ such that $Fail^M_t(x)$.  
\end{enumerate} 

Assume that $l$ and $f$ are the first failure and last success mutants found in $mut_K$, and $f'$ and $l'$ are the first failure and last success mutants found in $\bigcup_{n=1}^{K}mut_n$ by algorithm Alg. \ref{alg:ScenarioDomainAnalysis}

\vspace{+10pt}
\noindent{\textbf{\textit{Theorem \ref{thm:Expansion}.}}}

For any valid orderings of mutants in $mut_K$, there is a valid ordering of mutants in $\bigcup_{n=1}^{K}mut_n$ that preserves the ordering of the mutants in $mut_K$ such that  
\begin{enumerate}
\item  $f' = f$.
\item $l' = max\{l, l_o\}$, where 
\[l_o =\underset{x \in mut_{Old}}{\operatorname{argmax}}(x \preccurlyeq f), 
~~mut_{Old} = \bigcup_{n=1}^{K-1}mut_n.\] 
\end{enumerate}

\proof{Theorem \ref{thm:Expansion}} 

Let the set $mut_K$ be ordered in ascending order according to their distances to $t$ such that $mut_K=\{x_1,\cdots,x_k\}$, where $x_i \preccurlyeq x_j$ for $i<j=1, \cdots,k$. We also order the mutants in $mut_{Old}=\bigcup_{n=0}^{K-1}mut_n$ in ascending order, say $mut_{Old} = \{y_1, \cdots, y_{k'}\}$. We now construct the ordering of the set $mut_{All}=\bigcup_{n=0}^{K}mut_n$ in ascending order by inserting the elements in $mut_K$ into $mut_{Old}$ while preserving their orders in $mut_K$. This is feasible and the result is a valid ordering (i.e. in ascending order according to the distance to $t$) because, from assumption (1), we have that $mut_{Old}$ and $mut_K$ are disjoint. 

Assumption (2) implies that the first failure $f$ and last success $l$ exist in the set $mut_K$, and also $f'$ and $l'$ exist in $mut_{All}$. In the above ordering of all mutants in $\bigcup_{n=0}^{K}mut_n$, we can prove that the first failure mutant $f$ found in $mut_K$ must be the same as the first failure point $f'$ found in the set $\bigcup_{n=0}^K mut_n$. This is because, according to assumption (2), for all mutants $x$ in $\bigcup_{n=0}^{K-1}mut_n$, $\neg Fail^M_t(x)$. Therefore, for all $x \in mut_{All}$ we have that $Fail^M_t(x)$ implies that $x \in mut_K$. Since the ordering preserves the ordering of mutants in $mut_K$, we have that $f$ is the first failure in $mut_K$ implies that $f$ is also the first in $mut_{All}$ according to the above ordering. Therefore, $f = f'$, i.e. statement (1) of the theorem is true. 

Moreover, in above ordering of mutants, let $l_o$ be the mutant in $mut_{Old}$ listed just before $f' = f$. Formally, 
\[l_o =  \underset{x \in mut_{Old}}{\operatorname{argmax}}(x \preccurlyeq f).\] 

If $l \prec l_o$, $l_o$ must be listed after $l$ in the ordering of the set $mut_{All}$ because it is in the ascending order. Moreover, it must be last one before $f$ by its definitions. 

If $l_o \preccurlyeq l$, $l$ can be listed after $l_o$ because $l_o$ is not in $mut_K$, and it can be just before $f$. 

From the above two cases, we have that $l' = max\{l, l_o\}$. 

\qed

\vspace{+10pt}
\noindent{\textbf{\textit{Theorem \ref{thm:Ro}.}}}

If a set $Mut_t$ of $t$'s mutants is valid and relatively complete with a coverage size $\delta_t \geq \|t,FF(t)\|$, then, we have that $\|t,FF(t)\| = \Delta_{D^s}(t,M)$. 

\proof{Theorem \ref{thm:Ro}}

By the definition validity and the fact that $FF(t) \in Mut_t$, we have that 
\begin{equation}
FF(t) \in D^s. \label{eqn:thm1Proof1}
\end{equation}

By Definition \ref{def:MinimalDistanceToFailure}, because $Fail^M_t(FF(t))$, (\ref{eqn:thm1Proof1}) implies that 
\begin{equation}
\|t, FF(t)\| \geq \Delta_{D^s}(t,M). \label{eqn:thm1Proof2}
\end{equation}

By the definition of completeness, we have that 
\begin{equation}
\forall x \in D^s.\left( \|x,t\|\leq \delta_t \Rightarrow x \in Mut_t\right), \label{eqn:thm1Proof2b}
\end{equation}
where $\delta_t$ is the coverage size of $Mut_t$. 
Since $FF(t)$ is the first failure, by Lemma \ref{lmm:lmm1}, we have that 
\begin{equation}
\forall x \in Mut_t. \left(x \prec_t FF(t) \Rightarrow \neg Fail^M_t(x)\right). \label{eqn:thm1Proof3}
\end{equation}
When the coverage size $\delta_t \geq \|t,FF(t)\|$, (\ref{eqn:thm1Proof2b}) and (\ref{eqn:thm1Proof3}) imply that 
\begin{equation}
\forall x \in D^s.\left(x \prec_t FF(t) \Rightarrow \neg Fail^M_t(x)\right). \label{eqn:thm1Proof4}
\end{equation}
Therefore, by Definition \ref{def:MinimalDistanceToFailure}, we have that 
\begin{equation}
 \Delta_{D^s}(t,M) \geq \|t, FF(t)\|. \label{eqn:thm1Proof5}
\end{equation}
From (\ref{eqn:thm1Proof2}) and (\ref{eqn:thm1Proof5}), we have that $\|t,FF(t)\| = \Delta_{D^s}(t,M)$. 
\qed

\vspace{+10pt}
\noindent{\textbf{\textit{Theorem \ref{thm:R*}.}}}

If a set $Mut_t$ of $t$'s mutants is valid and relatively complete with a coverage size $\delta \geq \|t,FF(t)\|$, then, we have that $LS(t) \preccurlyeq_t \nabla_{D^s}(t,M)$, and there is no $x \in D^s$ such that $LS(t) \prec_t x \prec_t \nabla_{D^s}(t,M)$. 

\proof{Theorem \ref{thm:R*}}

We first prove that $LS(t) \preccurlyeq_t \nabla_{D^s}(t,M)$. 

By the definition of validity, we have that 
\begin{equation}
LS(t) \in D^s. \label{eqn:thm2Proof1}
\end{equation}
By the definition of $LS(t)$, we have that 
\begin{equation}
\neg Fail^M_x(LS(t)), \label{eqn:thm2Proof2}
\end{equation}
and
\begin{equation}
\forall x \in Mut_t.\left(x \prec_t LS(t) \Rightarrow \neg Fail^M_t(x)\right). \label{eqn:thm2Proof3}
\end{equation}
By the completeness of $Mut_t$,  we have that 
\begin{equation}
\forall x \in D^s.\left(\|x, t\| \leq \delta \Rightarrow x \in Mut_t \right). \label{eqn:thm2Proof4}
\end{equation}
where $\delta$ is the coverage size of $Mut_t$. 
From the fact that $\delta \geq \|t, LS(t)\|$, (\ref{eqn:thm2Proof3}) and (\ref{eqn:thm2Proof4}) imply that 
\begin{equation}
\forall x \in D^s. \left(x \prec_t LS(t) \Rightarrow \neg Fail^M_t(x)\right). \label{eqn:thm2Proof5}
\end{equation}

Definition \ref{def:SafeDistance}, (\ref{eqn:thm2Proof2}) and (\ref{eqn:thm2Proof5}) imply that $LS(t)$ is safe. Formally, 
\begin{equation}
Safe^M_x(LS(t)). \label{eqn:thm2Proof6a}
\end{equation}
Hence, by Lemma \ref{lmm:lmmD}, we have that
\begin{equation}
\|t,LS(t)\| \leq \nabla_{D^s}(t,M). \label{eqn:thm2Proof6}
\end{equation}

Now, we prove that there is no $x \in D^s$ such that $LS(t) \prec_t x \prec_t \nabla_{D^s}(t,M)$. 

By (\ref{eqn:thm2Proof6}), there are two possible situations: (a) $\|LS(t), t\| = \nabla_{D^s}(t,M)$, (b) $\|LS(t), t\| < \nabla_{D^s}(t,M)$. 

The theorem statement is trivially true in situation (a). 

In situation (b), 
by Definition \ref{def:SafeDistance} of $\nabla_{D^s}(t,M)$, the condition of the situation implies that 
there are $x \in D^s$ such that $x$ is safe and $LS(t) \prec_t x$. Formally, 
\begin{equation}
\exists x \in D^s.\left( Safe^M_t(x) \wedge (LS(t) \prec_t x ) \right).  \label{eqn:thm2Proof7}
\end{equation}

Let $\alpha$ be any element in $D^s$ that satisfies the predicate of (\ref{eqn:thm2Proof7}). By Lemma \ref{lmm:lmm3}, there is no $x \in Mut_t$ such that $LS(t) \prec_t x \prec_t FF(t)$. By the completeness of $Mut_t$ with a coverage size $\delta \geq \|t,FF(t)\|$, we have that there is no $x \in D^s$ such that $LS(t) \prec_t x \prec_t FF(t)$. Thus,  (\ref{eqn:thm2Proof7}) implies that 
\begin{equation}
FF(t) \preccurlyeq_t \alpha. \label{eqn:thm2Proof8}
\end{equation}
We now prove that it is impossible that $FF(t) \prec_t \alpha$. Otherwise, by the definition of $FF(t)$, we have that 
\begin{equation}
Fail^M_t(FF(t)). \label{eqn:thm2Proof9}
\end{equation}
By Definition \ref{def:SafeDistance}, (\ref{eqn:thm2Proof8}), (\ref{eqn:thm2Proof9}) would imply that $\alpha$ is not safe; formally 
\begin{equation}
FF(t) \prec_t \alpha \Rightarrow \neg Safe^M_t(\alpha).\label{eqn:thm2Proof9b}
\end{equation} 
This contradicts the assumption $Safe^M_t(\alpha)$. Thus, we have that $\alpha \preccurlyeq_t FF(t)$. 
Therefore, by Definition \ref{def:SafeDistance}, we have that
\begin{equation}
\left(\nabla_{D^s}(t,M)>\|t,LS(t)\|\right) \Rightarrow \left(\nabla_{D^s}(t,M) = \|t, FF(t)\|\right). \label{eqn:thm2Proof10b}
\end{equation}
By Lemma \ref{lmm:lmm3}, (\ref{eqn:thm2Proof10b}) implies that 
\begin{equation}
\neg \exists x \in Mut_t.\left( LS(t) \prec_t x \prec_t FF(t) \right). \label{eqn:thm2Proof11}
\end{equation}
By the validity and completeness of $Mut_t$ with coverage size $\delta \geq \|t, FF(t)\|$, we have that 
\begin{equation}
\forall x \in D.\left( \|x, t\| \leq \|FF(t),t\| \Rightarrow (x \in D^s \Leftrightarrow x \in Mut_t) \right). 
\label{eqn:thm2Proof12}
\end{equation}
(\ref{eqn:thm2Proof11}) and (\ref{eqn:thm2Proof12}) imply that
\begin{equation}
\neg \exists x \in D^s.\left( LS(t) \prec_t x \prec_t FS(t) \right). 
\end{equation}
Hence, theorem statement is true. 
\qed

\vspace{+10pt}
\noindent{\textbf{\textit{Theorem \ref{thm:Expansion2}.}}}

Assume that both $Mut_t$ and $Mut'_t$ are valid sets of the mutants of $t$, $\left<l, f \right>$ and $\left<l', f' \right>$ are the pairs of last success and first failure points found in $Mut_t$ and $Mut'_t$, respectively. Then, $Mut_t \subseteq Mut'_t$ implies that 

$l \preccurlyeq_t l'\preccurlyeq_t f' \preccurlyeq_t f,$ 
if $\|l, t\| \leq \nabla_{D^s}(t,M)$.\footnote{This condition normally holds when the set $Mut_t$ is generated with mutants closer to $t$ first.}  

\proof{Theorem \ref{thm:Expansion2}}

\noindent{\emph{(1) Proof of ($f' \preccurlyeq_t f$):}}

Because $f$ is the first failure of $Mut_t$, we have that $f \in Mut_t \subset Mut'_t$ and $Fail^M_t(f)$. By Lemma \ref{lmm:lmm1}, $f'$ is the first failure of $Mut'_t$. This implies that $f' \preccurlyeq_t f$. 

\noindent{\emph{(2) Proof of ($l \preccurlyeq_t l'$):}}

From the condition $\|l, t\| \leq \nabla_{D^s}(t,M)$ on $l$, we have that 
\begin{equation}
\forall x \in D^s.\left(x \prec_t l \Rightarrow Safe^M_t(x) \right). \label{eqn:Expansion2Proof1}
\end{equation}
By the validity of $Mut'_t$, (\ref{eqn:Expansion2Proof1}) implies that 
\begin{equation}
\forall x \in Mut'_t.\left(x \prec_t l \Rightarrow Safe^M_t(x) \right). \label{eqn:Expanions2Proof2}
\end{equation}
By Definition \ref{def:Safe}, (\ref{eqn:Expanions2Proof2}) implies that 
\begin{equation}
\forall x \in Mut'_t.\left(x \prec_t l \Rightarrow \neg Fail^M_t(x) \right).\label{eqn:Expansion2Proof3}
\end{equation}

Because $l'$ is the last success found in $Mut'_t$, $l'$ is the largest $\tau$ such that 
\begin{equation}
\neg Fail^M_t(\tau) \wedge \forall x \in Mut'_t.\left(x \prec_t \tau \Rightarrow \neg Fail^M_t(x) \right).\label{eqn:Expansion2Proof4}
\end{equation}
In other words, if $\tau$ satisfies (\ref{eqn:Expansion2Proof4}), then, $\tau \preccurlyeq l'$. 

Because $l$ is the last success found in $Mut_t$, we have that $\neg Fail^M_t(l)$ and $l \in Mut_t \subset Mut'_t$. Therefore, (\ref{eqn:Expansion2Proof3}) implies that $l$ satisfies the condition (\ref{eqn:Expansion2Proof4}). Thus, $l \preccurlyeq_t l'$. 
\qed


\newpage

\section*{Appendix B. Hypothesis Tests of Model Ranking}

We have conducted an inferential statistical analysis of the experiment data to rank the four models: GeminiPro, Codex, Falcon, and Llama2. All four models were evaluated on the benchmark containing 900 instances of code generation tasks with ten similarity metrics. All six unique model pairs were compared using paired two-tailed hypothesis $t$-tests. The Holm--Bonferroni correction is applied to control for multiple comparisons.

\subsection*{B.1. Design of The Experiment}

Given a distance metric $\| \cdot, \cdot \|$, on each coding task $i$ in the benchmark, the distance $\| i, x_i\|$ from $i$ to the first failure paraphrase $x_i$ found by model A was calculated, which is denoted by $ff_i^A$. Since the models were evaluated on the same tasks, paired $t$-tests were used to compare their scores on the tasks. For each pair of models A and B, the sample data set for a given similarity metric is, therefore, $\{ d_i = ff_i^A - ff_i^B \}$. We write $\mu_{A,B}$ as the mean of this sample data set, $s_{A,B}$ the standard deviation, and $n_{A,B}$ is the number of coding task instances with valid scores for both models A and B. 

The null and alternative hypotheses are defined as follows:

\begin{itemize}
    \item \textbf{Null Hypothesis} $H_0$: There is no difference between LLMs A and B in the $R^o$ robustness scores; formally, $\mu_{A,B} = 0$.
    \item \textbf{Alternative Hypothesis} $H_1$: LLMs A and B are different in the $R^o$ robustness scores; formally, $\mu_{A,B} \neq 0$.
\end{itemize}

The paired $t$-test statistic is calculated as follows:

\begin{equation}
    t_{A,B} = \frac{\bar{\mu_{A,B}}}{s_{A,B}/\sqrt{n_{A,B}}}
\end{equation}

\subsection*{B.2. Decision Rules}

We use a significance level of $\alpha = 0.05$ to assess differences between models. Since six pairwise comparisons are performed for each metric, the Holm--Bonferroni correction is applied separately within each similarity metric.

The statistical decisions are based on the adjusted $p$-values:

\begin{itemize}
    \item If the Holm-adjusted $p$-value is less than $0.05$, we reject $H_0$ and conclude that the two models have a statistically significant difference for that metric.
    \item If the Holm-adjusted $p$-value is greater than or equal to $0.05$, we fail to reject $H_0$. This indicates that the analysis does not provide sufficient evidence of a difference between the models. 
\end{itemize}

The sign of the mean difference, $\mu_{A,B}$, indicates which model has the higher numerical score. However, the better-performing model depends on the metric: higher values are generally preferred for similarity metrics, whereas lower values are preferred for distance metrics.

\subsection*{B.3. Results of The Hypothesis Tests}

Table~\ref{tab:hypothesis_test_results} presents the results of all 60 pairwise hypothesis tests across the ten evaluation metrics. The table reports mean difference, $t$-statistic, Holm-adjusted $p$-value, and statistical decision for each comparison.

Overall, 12 of the 60 comparisons are statistically significant after Holm--Bonferroni correction. These results indicate that significant differences occur for particular model pairs and metrics rather than uniformly across all comparisons. A discussion of the findings can be found in Section \ref{sec:Experiment}. 

\begin{table}[htbp]
\centering
\scriptsize
\caption{Results of Hypothesis Test ($\alpha = 0.05$, Holm-Bonferroni Adjusted)}
\label{tab:hypothesis_test_results}
\resizebox{8.5cm}{!}{%
\begin{tabular}{llrcrc}
\hline
\textbf{Metric} &
\textbf{Model Pair (A vs. B)} &
\textbf{Mean Diff. ($\bar{d}$)} &
\textbf{$t$-statistic} &
\textbf{$p$-value} &
\textbf{Decision} \\
\hline

\multirow{6}{*}{\textbf{BLEU}}
 & GeminiPro vs. Codex  & $-0.01170$ & $-1.29012$ & 0.592096 & No sig. diff. \\
 & GeminiPro vs. Falcon & $-0.01404$ & $-1.55202$ & 0.484138 & No sig. diff. \\
 & GeminiPro vs. Llama2 & $-0.01678$ & $-1.74769$ & 0.404419 & No sig. diff. \\
 & Codex vs. Falcon     & $-0.00227$ & $-3.59862$ & 0.002028 & Significant \\
 & Codex vs. Llama2     & $-0.00519$ & $-0.57502$ & 1.000000 & No sig. diff. \\
 & Falcon vs. Llama2    & $-0.00288$ & $-0.32026$ & 1.000000 & No sig. diff. \\
\hline

\multirow{6}{*}{\textbf{ChrF}}
 & GeminiPro vs. Codex  & $-1.13036$ & $-1.31377$ & 0.567846 & No sig. diff. \\
 & GeminiPro vs. Falcon & $-1.46679$ & $-1.70478$ & 0.372446 & No sig. diff. \\
 & GeminiPro vs. Llama2 & $-1.59481$ & $-1.78581$ & 0.372446 & No sig. diff. \\
 & Codex vs. Falcon     & $-0.26961$ & $-9.38836$ & $3.10 \times 10^{-19}$ & Significant \\
 & Codex vs. Llama2     & $-0.48347$ & $-0.55403$ & 1.000000 & No sig. diff. \\
 & Falcon vs. Llama2    & $-0.23763$ & $-0.27256$ & 1.000000 & No sig. diff. \\
\hline

\multirow{6}{*}{\textbf{Cosine}}
 & GeminiPro vs. Codex  & $-0.00098$ & $-0.22981$ & 1.000000 & No sig. diff. \\
 & GeminiPro vs. Falcon & $-0.00113$ & $-0.26729$ & 1.000000 & No sig. diff. \\
 & GeminiPro vs. Llama2 & $-0.00125$ & $-0.29528$ & 1.000000 & No sig. diff. \\
 & Codex vs. Falcon     & $-0.00016$ & $-6.83061$ & $9.47 \times 10^{-11}$ & Significant \\
 & Codex vs. Llama2     & $-0.00035$ & $-0.08405$ & 1.000000 & No sig. diff. \\
 & Falcon vs. Llama2    & $-0.00020$ & $-0.04834$ & 1.000000 & No sig. diff. \\
\hline

\multirow{6}{*}{\textbf{Euclidean}}
 & GeminiPro vs. Codex  & $0.023473$ & $1.122488$ & 0.797636 & No sig. diff. \\
 & GeminiPro vs. Falcon & $0.027162$ & $1.299523$ & 0.797636 & No sig. diff. \\
 & GeminiPro vs. Llama2 & $0.029280$ & $1.407890$ & 0.797636 & No sig. diff. \\
 & Codex vs. Falcon     & $0.003745$ & $7.75282$ & $1.49 \times 10^{-13}$ & Significant \\
 & Codex vs. Llama2     & $0.011056$ & $0.563328$ & 1.000000 & No sig. diff. \\
 & Falcon vs. Llama2    & $0.007275$ & $0.371260$ & 1.000000 & No sig. diff. \\
\hline

\multirow{6}{*}{\textbf{Levenshtein Char}}
 & GeminiPro vs. Codex  & $0.751962$ & $1.107858$ & 0.805026 & No sig. diff. \\
 & GeminiPro vs. Falcon & $1.012559$ & $1.490112$ & 0.546762 & No sig. diff. \\
 & GeminiPro vs. Llama2 & $2.054688$ & $2.914467$ & 0.018438 & Significant \\
 & Codex vs. Falcon     & $0.255599$ & $8.370772$ & $1.65 \times 10^{-15}$ & Significant \\
 & Codex vs. Llama2     & $0.466142$ & $0.665581$ & 1.000000 & No sig. diff. \\
 & Falcon vs. Llama2    & $0.192126$ & $0.274741$ & 1.000000 & No sig. diff. \\
\hline

\multirow{6}{*}{\textbf{Levenshtein Word}}
 & GeminiPro vs. Codex  & $0.008489$ & $0.487634$ & 1.000000 & No sig. diff. \\
 & GeminiPro vs. Falcon & $0.008489$ & $0.487634$ & 1.000000 & No sig. diff. \\
 & GeminiPro vs. Llama2 & $0.003378$ & $0.203959$ & 1.000000 & No sig. diff. \\
 & Codex vs. Falcon     & $0.000000$ & $0.000000$ & 1.000000 & No sig. diff. \\
 & Codex vs. Llama2     & $0.006861$ & $0.458521$ & 1.000000 & No sig. diff. \\
 & Falcon vs. Llama2    & $0.006861$ & $0.458521$ & 1.000000 & No sig. diff. \\
\hline

\multirow{6}{*}{\textbf{Meteor}}
 & GeminiPro vs. Codex  & $-0.01304$ & $-1.16578$ & 0.732094 & No sig. diff. \\
 & GeminiPro vs. Falcon & $-0.01565$ & $-1.39970$ & 0.647888 & No sig. diff. \\
 & GeminiPro vs. Llama2 & $-0.02109$ & $-1.88941$ & 0.295897 & No sig. diff. \\
 & Codex vs. Falcon     & $-0.00253$ & $-4.11673$ & 0.000252 & Significant \\
 & Codex vs. Llama2     & $-0.00632$ & $-0.58003$ & 1.000000 & No sig. diff. \\
 & Falcon vs. Llama2    & $-0.00371$ & $-0.34223$ & 1.000000 & No sig. diff. \\
\hline

\multirow{6}{*}{\textbf{Roberta}}
 & GeminiPro vs. Codex  & $-0.04928$ & $-3.44267$ & 0.001812 & Significant \\
 & GeminiPro vs. Falcon & $-0.05594$ & $-3.94299$ & 0.000348 & Significant \\
 & GeminiPro vs. Llama2 & $-0.07089$ & $-4.87142$ & $6.60 \times 10^{-6}$ & Significant \\
 & Codex vs. Falcon     & $-0.00661$ & $-5.76134$ & $6.93 \times 10^{-8}$ & Significant \\
 & Codex vs. Llama2     & $-0.01609$ & $-1.12006$ & 0.526007 & No sig. diff. \\
 & Falcon vs. Llama2    & $-0.00935$ & $-0.66030$ & 0.526007 & No sig. diff. \\
\hline

\multirow{6}{*}{\textbf{RougeL}}
 & GeminiPro vs. Codex  & $-0.01073$ & $-1.12639$ & 0.780959 & No sig. diff. \\
 & GeminiPro vs. Falcon & $-0.01255$ & $-1.32742$ & 0.738917 & No sig. diff. \\
 & GeminiPro vs. Llama2 & $-0.01448$ & $-1.47051$ & 0.708979 & No sig. diff. \\
 & Codex vs. Falcon     & $-0.00181$ & $-2.49241$ & 0.077236 & No sig. diff. \\
 & Codex vs. Llama2     & $-0.00558$ & $-0.59419$ & 1.000000 & No sig. diff. \\
 & Falcon vs. Llama2    & $-0.00372$ & $-0.39814$ & 1.000000 & No sig. diff. \\
\hline

\multirow{6}{*}{\textbf{RougeN}}
 & GeminiPro vs. Codex  & $-0.01366$ & $-1.24501$ & 0.640420 & No sig. diff. \\
 & GeminiPro vs. Falcon & $-0.01645$ & $-1.50646$ & 0.529291 & No sig. diff. \\
 & GeminiPro vs. Llama2 & $-0.02080$ & $-1.94840$ & 0.258493 & No sig. diff. \\
 & Codex vs. Falcon     & $-0.00270$ & $-3.14640$ & 0.010254 & Significant \\
 & Codex vs. Llama2     & $-0.00565$ & $-0.54003$ & 1.000000 & No sig. diff. \\
 & Falcon vs. Llama2    & $-0.00286$ & $-0.27523$ & 1.000000 & No sig. diff. \\
\hline
\end{tabular}%
}
\end{table}

\newpage

\section*{Appendix C. Statistical Analysis of The Margins of Errors}

As a part of the study on the statistical inference strength of observations made based on the experiment data, the margins of errors (MoE) were calculated with a 95\% confidence interval. For a sample data set of size $n$, the MoE was computed according to the following formula. 

$$
\mathrm{MoE} = Z \frac{s}{\sqrt{n}}
$$
where $Z = 1.96$ is the critical value for the 95\% confidence level, and $s$ is the standard deviation of the sample. 

\subsection*{C.1. MoE of Robustness Scores}

By definition of $R^o$ ($R^*$), the estimated $R^o$ ($R^*$) robustness score of a LLM $A$ evaluated with a similarity metric $\|\cdot,\cdot\|$ is the mean values of the sample set $\{ff_x | x \in D\}$, where $D$ is the benchmark of coding tasks. For each $x \in D$, $ff_x$ is the distances from the original coding task $x$ to the first failure (last success) paraphrase $x'$ of $x$ found by model $A$, i.e. $ff_x=\|x,x'\|$. To quantify the precision of the estimation, the margin of error (MoE) of the mean value of the sample set $\{ff_x | x \in D\}$ is calculated with $Z$'s value corresponding to a 95\% confidence interval. The experiment has been conducted with a benchmark of 900 coding tasks; see Section \ref{sec:Experiment}. Thus, the sample size $n = 900$. 

In Table~\ref{tab:combined_robustness} in Section \ref{sec:Experiment}, the robustness scores are presented together with corresponding the MoE in the form of $x \pm e$, where $x$ denotes the estimated mean value of the samples, i.e. the $R^o$ ($R^*$), and $e$ is the corresponding MoE.

Table \ref{tab:RelativeMoE} below gives the relative MoE (i.e. $\frac{e}{x}$, denoted by MoE\%) for each robustness score. It clearly shows that the relative MoEs are all very small. The maximal is 5.2803\% and the overall average is only 1.908\%. This indicates that the estimates of $R^o$ and $R^*$ are of very high precision.

\begin{table*}[hb]
\caption{Relative Margin of Error in the Estimated Robustness Scores}
\label{tab:RelativeMoE}
\includegraphics[width=\textwidth]{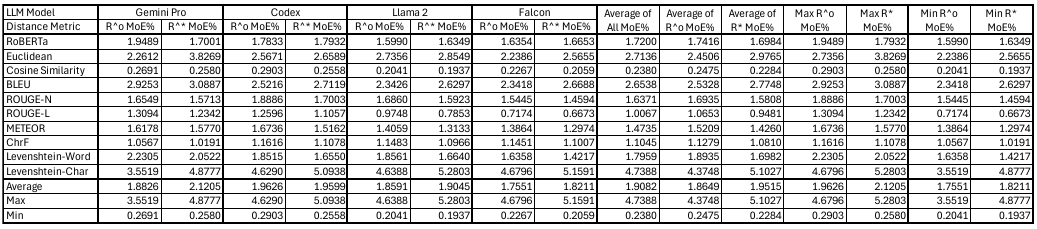}
\end{table*}


\subsection*{C.2. MoE of The Average Values of $n$ and $k$}
In the analysis of the tipping points, the MoE values are presented in Table~\ref{tab:averageNandK}. Their relative MoEs are given below in Table \ref{tab:RelativeMoE_TippingPints}. The data shows that the relative MoEs of the average value of $n$ and $k$ are all very small. The maximal MoE\% for $n$ is 5.8366\%, while the average of MoE\%s for $n$ is 5.0035\%. For the relative MoEs of the average value of $k$, the maximal is only 2.9557\%, while the average is 1.6267. This indicates that the estimations of the average values of $n$ and $k$ for different models and using different metrics are of high precision. 

\begin{table*}[htbp]
\centering
\caption{Relative MoE of the Average Values of ($n$ and $k$) of the Tipping Points }
\label{tab:RelativeMoE_TippingPints}
\resizebox{\textwidth}{!}{
\begin{tabular}{|l|c|c|c|c|c|c|c|c||c|c|c|c|c|c|}
\hline
\multirow{2}{*}{\textbf{Metric}} & \multicolumn{2}{c|}{\textbf{Codex}} & \multicolumn{2}{c|}{\textbf{GeminiPro}} & \multicolumn{2}{c|}{\textbf{Llama2}} & \multicolumn{2}{c||}{\textbf{Falcon}} &\multicolumn{2}{c|}{\textbf{Average}} & \multicolumn{2}{c|}{\textbf{Max}} & \multicolumn{2}{c|}{\textbf{Min}}\\ \cline{2-15} 
 & $n ~ \text{MoE}\%$ & $k ~ \text{MoE}\%$ & $n ~ \text{MoE}\%$ & $k ~ \text{MoE}\%$ & $n ~ \text{MoE}\%$ & $k ~ \text{MoE}\%$ & $n ~ \text{MoE}\%$ & $k ~ \text{MoE}\%$ & $n ~ \text{MoE}\%$ & $k ~ \text{MoE}\%$ & $n ~ \text{MoE}\%$ & $k ~ \text{MoE}\%$ & $n ~ \text{MoE}\%$ & $k ~ \text{MoE}\%$\\ \hline
RoBERTa &3.8147&1.9704&4.1451&1.8315&4.5283&1.6461&4.8689&2.9557&4.3393&2.1009&4.8689&2.9557&3.8147&1.6461\\ \hline
ChrF &5.2434&1.9084&5.5970&1.9704&4.4715&1.4218&4.5455&1.8018&4.9644&1.7756&5.5970&1.9704&4.4715&1.4218\\ \hline
Euclidean &5.4348&1.8315&5.2434&1.6129&4.8583&1.7241&4.8507&1.9704&5.0968&1.7847&5.4348&1.9704&4.8507&1.6129\\ \hline
Cosine &5.5944&1.7937&5.2434&1.7167&4.8387&1.6598&4.6218&1.7167&5.0746&1.7217&5.5944&1.7937&4.6218&1.6598\\ \hline
BLEU &5.3232&1.9139&5.4054&1.9011&4.9430&1.4778&4.9430&1.9608&5.1536&1.8134&5.4054&1.9608&4.9430&1.4778\\ \hline
METEOR &5.5970&1.9704&5.3061&1.6461&4.6512&1.5686&4.9618&1.6807&5.1290&1.7165&5.5970&1.9704&4.6512&1.5686\\ \hline
Levenshtein-Char &5.8366&1.9704&4.6875&1.9704&5.0584&1.4778&5.0584&1.9704&5.1602&1.8473&5.8366&1.9704&4.6875&1.4778\\ \hline
Levenshtein-Word &5.0000&2.1127&4.7753&2.2556&4.4715&1.8349&4.9587&1.7544&4.8014&1.9894&5.0000&2.2556&4.4715&1.7544\\ \hline
ROUGE-N &5.4152&1.8182&5.3512&1.8519&4.4715&1.4218&4.7210&1.8868&4.9897&1.7447&5.4152&1.8868&4.4715&1.4218\\ \hline
ROUGE-L &5.0584&1.7241&5.4152&1.7094&4.5643&1.4019&4.5249&1.4778&4.8907&1.5783&5.4152&1.7241&4.5249&1.4019\\ \hline\hline
\textbf{Average} &5.2083&1.8433&5.2817&1.7778&4.7619&1.4151&4.7619&1.4706&5.0035&1.6267&5.2817&1.8433&4.7619&1.4151\\ \hline
\textbf{Max} &5.8366&2.1127&5.5970&2.2556&5.0584&1.8349&5.0584&2.9557&5.1602&2.1009&5.8366&2.9557&5.0584&1.8349\\ \hline
\textbf{Min} &3.8147&1.7241&4.1451&1.6129&4.4715&1.4019&4.5249&1.4778&4.3393&1.5783&4.5249&1.7241&3.8147&1.4019\\ \hline
\end{tabular}
}
\end{table*}

\subsection*{C.3. MoE of The Average Number of LLM Queries}

The average numbers of LLM queries in the evaluation of a LLM's robustness with the technique proposed in this paper have been presented in Table~\ref{tab:AverageLLMQueries} together with their corresponding MoEs. The MoEs are relatively small compared with the corresponding average number of LLM queries as shown in Table \ref{tab:MoE_QueryBudget_MoE}. The maximal relative MoE is 12.9993\%, while the average is 12.6084\%, indicating that the estimated query requirements for reaching the first failure are reasonable precise. The absolute value of MoE ranges from 1.71 to 1.98 across the similarity metrics and LLMs. Overall, the relatively small MoEs suggest limited uncertainty in the reported average number of paraphrases required to reach the first failure.

\begin{table}[htbp]
\centering
\caption{Relative MoE of the Average Numbers of LLM Queries Per Test Case}
\label{tab:MoE_QueryBudget_MoE}
\resizebox{\linewidth}{!}{
\begin{tabular}{|l|c|c|c|c|c|c|c|}
\hline
\textbf{Metric} & \textbf{GeminiPro} & \textbf{Codex} & \textbf{Falcon} & \textbf{Llama2} & \textbf{Average} & \textbf{Min} & \textbf{Max}\\ \hline
BLEU             &12.4831&12.8804&12.9079&12.8205&12.7730&12.4831&12.9079\\ \hline
ChrF             &12.1099&12.4511&12.4916&12.3906&12.3608&12.1099&12.4916\\ \hline
Levenshtein-Char &11.9062&12.2180&12.2108&12.1762&12.1278&11.9062&12.2180\\ \hline
Levenshtein-Word &12.4833&12.8223&12.8132&12.6893&12.7020&12.4833&12.8223\\ \hline
ROUGE-N          &12.3772&12.7485&12.7196&12.7131&12.6396&12.3772&12.7485\\ \hline
ROUGE-L          &12.2898&12.8028&12.7750&12.6973&12.6412&12.2898&12.8028\\ \hline
METEOR           &12.5952&12.9993&12.9657&12.8668&12.8567&12.5952&12.9993\\ \hline
RoBERTa          &12.5676&12.8369&12.8634&12.7753&12.7608&12.5676&12.8634\\ \hline
Cosine           &12.5417&12.7793&12.7949&12.7076&12.7059&12.5417&12.7949\\ \hline
Euclidean        &12.5503&12.7809&12.7965&12.6996&12.7068&12.5503&12.7965\\ \hline\hline
\textbf{Average} &12.4097&12.7323&12.7029&12.5889&12.6084&12.4097&12.7323\\ \hline
\textbf{Min}     &11.9062&12.2180&12.2108&12.1762&12.1278&11.9062&12.2180\\ \hline
\textbf{Max}     &12.5952&12.9993&12.9657&12.8668&12.8567&12.5952&12.9993\\ \hline
\end{tabular}
}
\end{table}

\subsection*{C.4. MoE of The Average Times to Calculate the Distances}

The MoE values of the average times spent on claculating the distances between the original coding task and its paraphrases have been presented in Table~\ref{tab:distanceMetricTime}. The 95\% MoE ranges from 0.0145 to 0.0354 ms across the metrics. The maximal relative MoE is 6.3777\%, while the average is 2.6140\%. They are very small relative to the corresponding average computation times, indicating that the estimated times required to calculate the distance between the original task description and its paraphrase are of high precision.

\begin{table}[htbp]
\centering
\caption{Times to Calculate the Distance bwt The Original Task Description And
A Paraphrase}
\label{tab:Table8_ComputationTimes}
\resizebox{\linewidth}{!}{
\begin{tabular}{|l|c|c|c|c|c|}
\hline
\textbf{Metric} & \textbf{Average (ms)} & \textbf{95\% MoE ($\pm$)} & \textbf{StDev ($s$)} & \textbf{Max} & \textbf{Min} \\ \hline
BLEU             & 0.2399 & 0.0153 & 0.0078 & 0.2922 & 0.2304 \\ \hline
ChrF             & 0.2489 & 0.0152 & 0.0078 & 0.2970 & 0.2394 \\ \hline
Levenshtein-Char & 0.4730 & 0.0164 & 0.0079 & 0.5229 & 0.4634 \\ \hline
Levenshtein-Word & 0.4799 & 0.0145 & 0.0079 & 0.5248 & 0.4704 \\ \hline
RougeN           & 0.8812 & 0.0161 & 0.0082 & 0.9333 & 0.8714 \\ \hline
RougeL           & 0.9119 & 0.0153 & 0.0078 & 0.9629 & 0.9024 \\ \hline
Meteor           & 1.1671 & 0.0161 & 0.0082 & 1.2213 & 1.1575 \\ \hline
RoBERTa          & 3.6302 & 0.0261 & 0.0082 & 3.6795 & 3.6205 \\ \hline
Cosine           & 3.7341 & 0.0251 & 0.0077 & 3.7751 & 3.7244 \\ \hline
Euclidean        & 3.9699 & 0.0354 & 0.0077 & 4.0172 & 3.9604 \\ \hline
\end{tabular}
}
\end{table}

\end{document}